\documentclass[amsmath,amssymb,aps,10pt,prd,letterpaper,nofootinbib,balancelastpage,notitlepage,superscriptaddress,twocolumn,floatfix]{revtex4-2}
\usepackage{graphicx}	% Standard figures 
\usepackage{epstopdf}	% Convert eps to pdf
\usepackage[sort&compress]{natbib}	 	% Standard reference package
	\setcitestyle{square,numbers,comma}	% Set citation style
\usepackage[colorlinks=true,urlcolor=blue,linkcolor=black,citecolor=red]{hyperref}
\usepackage{verbatim}

\newcommand{\avg}[1]{\left \langle #1 \right \rangle} % expectation value
\newcommand{\eq}[1]{\begin{equation} #1 \end{equation}} % equation environment
\newcommand{\spliteq}[1]{\begin{equation} \begin{split} #1 \end{split} \end{equation}} % equation environment with split
\newcommand{\aligneq}[1]{\begin{align} #1 \end{align}} % equation with align
\newcommand{\parens}[1]{\left ( #1 \right )} % parentheses
\newcommand{\brackets}[1]{\left [ #1 \right ]} % brackets
\newcommand{\squiggles}[1]{\left \{ #1 \right \}} % braces
\newcommand{\eqrefTemp}[1]{Eq.~\ref{#1}}
\renewcommand{\eqref}{\eqrefTemp}
\newcommand{\secref}[1]{Sec.~\ref{#1}}
\newcommand{\figref}[1]{Fig.~\ref{#1}}

\newcommand{\tabref}[1]{Tab.~\ref{#1}}

\newcommand{\unitVec}[1]{\mathbf{\hat{#1}}} % unit vector hat/bold notation
\newcommand{\var}{\operatorname{Var}}
\newcommand{\cov}[2]{\operatorname{Cov}\brackets{#1,#2}}

\newcommand{\Qavg}[1]{\avg{Q^{#1}(\unitVec{n})}}
\newcommand{\Uavg}[1]{\avg{U^{#1}(\unitVec{n})}}
\newcommand{\QavgCMB}{\Qavg{(\mathrm{CMB})}}

\newcommand{\UavgCMB}{\Uavg{(\mathrm{CMB})}}

\newcommand{\cit}[1]{c_i^{#1}(\unitVec{n},t)}
\newcommand{\citau}[1]{\bar{c}_i^{#1}(\unitVec{n},\tau)}
\newcommand{\sit}[1]{s_i^{#1}(\unitVec{n},t)}
\newcommand{\sitau}[1]{\bar{s}_i^{#1}(\unitVec{n},\tau)}
\newcommand{\ftau}{\bar{f}(\tau)}
\newcommand{\Dit}[1]{D_i^{#1}(\unitVec{n},t)}
\newcommand{\Distatict}{\Dit{(\mathrm{static})}}
\newcommand{\Diosct}{\Dit{(\mathrm{osc})}}
\newcommand{\Dinoiset}{\Dit{(\mathrm{N})}}
\newcommand{\Difgt}{\Dit{(\mathrm{fg})}}
\newcommand{\Dicmbt}{\Dit{(\mathrm{CMB})}}
\newcommand{\Dircmbt}{\Dit{(\mathrm{rCMB})}}
\newcommand{\Ditau}[1]{\bar{D}_i^{#1}(\unitVec{n},\tau)}

\newcommand{\Diosctau}{\Ditau{(\mathrm{osc})}}
\newcommand{\Difgtau}{\Ditau{(\mathrm{fg})}}
\newcommand{\Dicmbtau}{\Ditau{(\mathrm{CMB})}}
\newcommand{\Dinoisetau}{\Ditau{(\mathrm{N})}}
\newcommand{\Dircmbtau}{\Ditau{(\mathrm{rCMB})}}
\newcommand{\DisDtau}{\Ditau{(s_D)}}
\newcommand{\rit}[1]{r_i^{#1}(\unitVec{n},t)}
\newcommand{\ricmbt}{\rit{(\mathrm{CMB})}}
\newcommand{\ritau}[1]{\bar{r}_i^{#1}(\unitVec{n},\tau)}
\newcommand{\ricmbtau}{\ritau{(\mathrm{CMB})}}
\newcommand{\rifgtau}{\ritau{(\mathrm{fg})}}
\newcommand{\rinoisetau}{\ritau{(\mathrm{N})}}

\newcommand{\risrtau}{\ritau{(s_r)}}
\newcommand{\risronetau}{\ritau{(s_{r_1})}}
\newcommand{\risrtwotau}{\ritau{(s_{r_2})}}
\newcommand{\Qcoadd}[1]{\bar{Q}^{#1}(\unitVec{n})}
\newcommand{\Ucoadd}[1]{\bar{U}^{#1}(\unitVec{n})}
\newcommand{\QcoaddCMB}{\Qcoadd{(\mathrm{CMB})}}

\newcommand{\Qcoaddfg}{\Qcoadd{(\mathrm{fg})}}

\newcommand{\Qcoaddnoise}{\Qcoadd{(\mathrm{N})}}
\newcommand{\Qcoaddosc}{\Qcoadd{(\mathrm{osc})}}

\newcommand{\witau}[1]{w_i^{#1}(\unitVec{n},\tau)}
\newcommand{\vitau}[1]{v_i^{#1}(\unitVec{n},\tau)}
\newcommand{\rhoel}[2]{\rho^{(#1,#2)}(\tau)}
\newcommand{\rhoiel}[2]{\rho_i^{(#1,#2)}(\tau)}
\newcommand{\rhosrsD}{\rhoel{s_r}{s_D}}
\newcommand{\rhoisrsD}{\rhoiel{s_r}{s_D}}

\newcommand{\rhocmbnoise}{\rhoel{\mathrm{CMB}}{\mathrm{N}}}

\newcommand{\rhonoisenoise}{\rhoel{\mathrm{N}}{\mathrm{N}}}

\newcommand{\rhocmbosc}{\rhoel{\mathrm{CMB}}{\mathrm{osc}}}
\newcommand{\rhosrosc}{\rhoel{s_r}{\mathrm{osc}}}
\newcommand{\rhosrrcmb}{\rhoel{s_r}{\mathrm{rCMB}}}
\newcommand{\Rel}[2]{R^{(#1,#2)}(\tau)}

\newcommand{\Riel}[2]{R_i^{(#1,#2)}(\tau)}
\newcommand{\Rsrsr}{\Rel{s_{r_1}}{s_{r_2}}}

\newcommand{\Risrsr}{\Riel{s_{r_1}}{s_{r_2}}}
\newcommand{\Rcmbcmb}{\Rel{\mathrm{CMB}}{\mathrm{CMB}}}

\newcommand{\Rnoisenoise}{\Rel{\mathrm{N}}{\mathrm{N}}}

\newcommand{\Rfgfg}{\Rel{\mathrm{fg}}{\mathrm{fg}}}
\newcommand{\fhatel}[2]{\hat{f}^{(#1,#2)}(\tau)}
\newcommand{\fhatsrsD}{\fhatel{s_r}{s_D}}
\newcommand{\fhatsrosc}{\fhatel{s_r}{\mathrm{osc}}}
\newcommand{\fhatsrrcmb}{\fhatel{s_r}{\mathrm{rCMB}}}

\newcommand{\fhattype}[1]{\hat{f}^{(#1)}(\tau)}
\newcommand{\fhatosc}{\fhattype{\mathrm{osc}}}
\newcommand{\fhatdyn}{\fhattype{\mathrm{dyn}}}
\newcommand{\fhatbkg}{\fhattype{\mathrm{bkg}}}
\newcommand{\fhatavg}{\avg{ \hat{f}(\tau) }}
\newcommand{\fhatbkgavg}{\avg{ \fhatbkg }}
\newcommand{\fhatdynavg}{\avg{ \fhatdyn }}

\newcommand{\fjk}{\hat{f}^{(\mathrm{jk})}(\tau)}
\newcommand{\deltaqmjk}{\Delta q_m^{(\mathrm{jk})}}
\newcommand{\Ajk}{\hat{A}_m^{(\mathrm{jk})}}
\newcommand{\Adc}{\hat{A}_0^{(\mathrm{jk})}}
\newcommand{\ajk}{\hat{a}^{(\mathrm{jk})}}
\newcommand{\deltaqjk}{\Delta\hat{q}^{(\mathrm{jk})}}

\begin{document}

%\preprint{}

\title{BICEP / \emph{Keck} XII: Constraints on axion-like polarization oscillations in the cosmic microwave background}


%\author{}
%\thanks{}
%\affiliation{}

\author{BICEP/\emph{Keck} Collaboration: P.~A.~R.~Ade}
\affiliation{School of Physics and Astronomy, Cardiff University,
  Cardiff, CF24 3AA, United Kingdom}
\author{Z.~Ahmed}
\affiliation{Kavli Institute for Particle Astrophysics and Cosmology,
  SLAC National Accelerator Laboratory, 2575 Sand Hill Rd, Menlo Park,
  CA 94025, USA}
\author{M.~Amiri}
\affiliation{Department of Physics and Astronomy, University of
  British Columbia, Vancouver, British Columbia, V6T 1Z1, Canada}
\author{D.~Barkats}
\affiliation{Harvard-Smithsonian Center for Astrophysics, 60 Garden
  Street MS 42, Cambridge, Massachusetts 02138, USA}
\author{R.~Basu Thakur}
\affiliation{Department of Physics, California Institute of
  Technology, Pasadena, California 91125, USA}
\author{C.~A.~Bischoff}
\affiliation{Department of Physics, University of Cincinnati,
  Cincinnati, Ohio 45221, USA}
\author{J.~J.~Bock}
\affiliation{Department of Physics, California Institute of
  Technology, Pasadena, California 91125, USA}
\affiliation{Jet Propulsion Laboratory, Pasadena, California 91109,
  USA}
\author{H.~Boenish}
\affiliation{Harvard-Smithsonian Center for Astrophysics, 60 Garden
  Street MS 42, Cambridge, Massachusetts 02138, USA}
\author{E.~Bullock}
\affiliation{Minnesota Institute for Astrophysics, University of
  Minnesota, Minneapolis, Minnesota 55455, USA}
\author{V.~Buza}
\affiliation{Kavli Institute for Cosmological Physics, University of
  Chicago, Chicago, IL 60637, USA}
\author{J.~R.~Cheshire~IV}
\affiliation{Minnesota Institute for Astrophysics, University of
  Minnesota, Minneapolis, Minnesota 55455, USA}
\author{J.~Connors}
\affiliation{Harvard-Smithsonian Center for Astrophysics, 60 Garden
  Street MS 42, Cambridge, Massachusetts 02138, USA}
\affiliation{National Institute of Standards and Technology, Boulder,
  Colorado 80305, USA}
\author{J.~Cornelison}
\affiliation{Harvard-Smithsonian Center for Astrophysics, 60 Garden
  Street MS 42, Cambridge, Massachusetts 02138, USA}
\author{M.~Crumrine}
\affiliation{School of Physics and Astronomy, University of Minnesota,
  Minneapolis, Minnesota 55455, USA}
\author{A.~Cukierman}
\email[Corresponding author: A.~Cukierman\\]{ajcukier@stanford.edu}
\affiliation{Kavli Institute for Particle Astrophysics and Cosmology,
  SLAC National Accelerator Laboratory, 2575 Sand Hill Rd, Menlo Park,
  CA 94025, USA}
\affiliation{Department of Physics, Stanford University, Stanford, CA
  94305, USA}
\author{M.~Dierickx}
\affiliation{Harvard-Smithsonian Center for Astrophysics, 60 Garden
  Street MS 42, Cambridge, Massachusetts 02138, USA}
\author{L.~Duband}
\affiliation{Service des Basses Temp\'{e}ratures, Commissariat \`{a}
  l'Energie Atomique, 38054 Grenoble, France}
\author{S.~Fatigoni}
\affiliation{Department of Physics and Astronomy, University of
  British Columbia, Vancouver, British Columbia, V6T 1Z1, Canada}
\author{J.~P.~Filippini}
\affiliation{Department of Physics, University of Illinois at
  Urbana-Champaign, Urbana, Illinois 61801, USA}
\affiliation{Department of Astronomy, University of Illinois at
  Urbana-Champaign, Urbana, Illinois 61801, USA}
\affiliation{Department of Physics, California Institute of
  Technology, Pasadena, California 91125, USA}
\author{S.~Fliescher}
\affiliation{School of Physics and Astronomy, University of Minnesota,
  Minneapolis, Minnesota 55455, USA}
\author{N.~Goeckner-Wald}
\affiliation{Department of Physics, Stanford University, Stanford, CA
  94305, USA}
\author{J.~Grayson}
\affiliation{Department of Physics, Stanford University, Stanford, CA
  94305, USA}
\author{G.~Hall}
\affiliation{School of Physics and Astronomy, University of Minnesota,
  Minneapolis, Minnesota 55455, USA}
\affiliation{Department of Physics, Stanford University, Stanford, CA
  94305, USA}
\author{M.~Halpern}
\affiliation{Department of Physics and Astronomy, University of
  British Columbia, Vancouver, British Columbia, V6T 1Z1, Canada}
\author{S.~Harrison}
\affiliation{Harvard-Smithsonian Center for Astrophysics, 60 Garden
  Street MS 42, Cambridge, Massachusetts 02138, USA}
\author{S.~Henderson}
\affiliation{Kavli Institute for Particle Astrophysics and Cosmology,
  SLAC National Accelerator Laboratory, 2575 Sand Hill Rd, Menlo Park,
  CA 94025, USA}
\author{S.~R.~Hildebrandt}
\affiliation{Jet Propulsion Laboratory, Pasadena, California 91109,
  USA}
\affiliation{Department of Physics, California Institute of
  Technology, Pasadena, California 91125, USA}
\author{G.~C.~Hilton}
\affiliation{National Institute of Standards and Technology, Boulder,
  Colorado 80305, USA}
\author{J.~Hubmayr}
\affiliation{National Institute of Standards and Technology, Boulder,
  Colorado 80305, USA}
\author{H.~Hui}
\affiliation{Department of Physics, California Institute of
  Technology, Pasadena, California 91125, USA}
\author{K.~D.~Irwin}
\affiliation{Department of Physics, Stanford University, Stanford, CA
  94305, USA}
\author{J.~Kang}
\affiliation{Department of Physics, Stanford University, Stanford, CA
  94305, USA}
\author{K.~S.~Karkare}
\affiliation{Kavli Institute for Cosmological Physics, University of
  Chicago, Chicago, IL 60637, USA}
\author{E.~Karpel}
\affiliation{Department of Physics, Stanford University, Stanford, CA
  94305, USA}
\author{B.~G.~Keating}
\affiliation{Department of Physics, University of California at San
  Diego, La Jolla, California 92093, USA}
\author{S.~Kefeli}
\affiliation{Department of Physics, California Institute of
  Technology, Pasadena, California 91125, USA}
\author{S.~A.~Kernasovskiy}
\affiliation{Department of Physics, Stanford University, Stanford, CA
  94305, USA}
\author{J.~M.~Kovac}
\affiliation{Harvard-Smithsonian Center for Astrophysics, 60 Garden
  Street MS 42, Cambridge, Massachusetts 02138, USA}
\affiliation{Department of Physics, Harvard University, Cambridge, MA
  02138, USA}
\author{C.~L.~Kuo}
\affiliation{Department of Physics, Stanford University, Stanford, CA
  94305, USA}
\affiliation{Kavli Institute for Particle Astrophysics and Cosmology,
  SLAC National Accelerator Laboratory, 2575 Sand Hill Rd, Menlo Park,
  CA 94025, USA}
\author{K.~Lau}
\affiliation{School of Physics and Astronomy, University of Minnesota,
  Minneapolis, Minnesota 55455, USA}
\author{E.~M.~Leitch}
\affiliation{Kavli Institute for Cosmological Physics, University of
  Chicago, Chicago, IL 60637, USA}
\author{K.~G.~Megerian}
\affiliation{Jet Propulsion Laboratory, Pasadena, California 91109,
  USA}
\author{L.~Moncelsi}
\affiliation{Department of Physics, California Institute of
  Technology, Pasadena, California 91125, USA}
\author{T.~Namikawa}
\affiliation{Department of Physics, Stanford University, Stanford, CA
  94305, USA}
\author{C.~B.~Netterfield}
\affiliation{Department of Physics, University of Toronto, Toronto,
  Ontario, M5S 1A7, Canada}
\author{H.~T.~Nguyen}
\affiliation{Jet Propulsion Laboratory, Pasadena, California 91109,
  USA}
\affiliation{Department of Physics, California Institute of
  Technology, Pasadena, California 91125, USA}
\author{R.~O'Brient}
\affiliation{Jet Propulsion Laboratory, Pasadena, California 91109,
  USA}
\affiliation{Department of Physics, California Institute of
  Technology, Pasadena, California 91125, USA}
\author{R.~W.~Ogburn~IV}
\affiliation{Department of Physics, Stanford University, Stanford, CA
  94305, USA}
\author{S.~Palladino}
\affiliation{Department of Physics, University of Cincinnati,
  Cincinnati, Ohio 45221, USA}
\author{T.~Prouve}
\affiliation{Service des Basses Temp\'{e}ratures, Commissariat \`{a}
  l'Energie Atomique, 38054 Grenoble, France}
\author{C.~Pryke}
\affiliation{School of Physics and Astronomy, University of Minnesota,
  Minneapolis, Minnesota 55455, USA}
\affiliation{Minnesota Institute for Astrophysics, University of
  Minnesota, Minneapolis, Minnesota 55455, USA}
\author{B.~Racine}
\affiliation{Harvard-Smithsonian Center for Astrophysics, 60 Garden
  Street MS 42, Cambridge, Massachusetts 02138, USA}
\affiliation{Aix-Marseille Universit\'{e}, CNRS/IN2P3, CPPM, Marseille, France}
 \author{C.~D.~Reintsema}
 \affiliation{National Institute of Standards and Technology, Boulder,
  Colorado 80305, USA}
\author{S.~Richter}
\affiliation{Harvard-Smithsonian Center for Astrophysics, 60 Garden
  Street MS 42, Cambridge, Massachusetts 02138, USA}
\author{A.~Schillaci}
\affiliation{Department of Physics, California Institute of
  Technology, Pasadena, California 91125, USA}
\author{B.~L.~Schmitt}
\affiliation{Harvard-Smithsonian Center for Astrophysics, 60 Garden
  Street MS 42, Cambridge, Massachusetts 02138, USA}
\author{R.~Schwarz}
\affiliation{School of Physics and Astronomy, University of Minnesota,
  Minneapolis, Minnesota 55455, USA}
\author{C.~D.~Sheehy}
\affiliation{School of Physics and Astronomy, University of Minnesota,
  Minneapolis, Minnesota 55455, USA}
\affiliation{Kavli Institute for Cosmological Physics, University of
  Chicago, Chicago, IL 60637, USA}
\author{A.~Soliman}
\affiliation{Department of Physics, California Institute of
  Technology, Pasadena, California 91125, USA}
\author{T.~St.~Germaine}
\affiliation{Harvard-Smithsonian Center for Astrophysics, 60 Garden
  Street MS 42, Cambridge, Massachusetts 02138, USA}
\author{B.~Steinbach}
\affiliation{Department of Physics, California Institute of
  Technology, Pasadena, California 91125, USA}
\author{R.~V.~Sudiwala}
\affiliation{School of Physics and Astronomy, Cardiff University,
  Cardiff, CF24 3AA, United Kingdom}
\author{G.~Teply}
\affiliation{Department of Physics, California Institute of
  Technology, Pasadena, California 91125, USA}
\author{K.~L.~Thompson}
\affiliation{Department of Physics, Stanford University, Stanford, CA
  94305, USA}
\affiliation{Kavli Institute for Particle Astrophysics and Cosmology,
  SLAC National Accelerator Laboratory, 2575 Sand Hill Rd, Menlo Park,
  CA 94025, USA}
\author{J.~E.~Tolan}
\affiliation{Department of Physics, Stanford University, Stanford, CA
  94305, USA}
\author{C.~Tucker}
\affiliation{School of Physics and Astronomy, Cardiff University,
  Cardiff, CF24 3AA, United Kingdom}
\author{A.~D.~Turner}
\affiliation{Jet Propulsion Laboratory, Pasadena, California 91109,
  USA}
\author{C.~Umilta}
\affiliation{Department of Physics, University of Illinois at
  Urbana-Champaign, Urbana, Illinois 61801, USA}
\author{A.~G.~Vieregg}
\affiliation{Kavli Institute for Cosmological Physics, University of
  Chicago, Chicago, IL 60637, USA}
\affiliation{Department of Physics, Enrico Fermi Institute, University
  of Chicago, Chicago, IL 60637, USA}
\author{A.~Wandui}
\affiliation{Department of Physics, California Institute of
  Technology, Pasadena, California 91125, USA}
\author{A.~C.~Weber}
\affiliation{Jet Propulsion Laboratory, Pasadena, California 91109,
  USA}
\author{D.~V.~Wiebe}
\affiliation{Department of Physics and Astronomy, University of
  British Columbia, Vancouver, British Columbia, V6T 1Z1, Canada}
\author{J.~Willmert}
\affiliation{School of Physics and Astronomy, University of Minnesota,
  Minneapolis, Minnesota 55455, USA}
\author{C.~L.~Wong}
\affiliation{Harvard-Smithsonian Center for Astrophysics, 60 Garden
  Street MS 42, Cambridge, Massachusetts 02138, USA}
\author{W.~L.~K.~Wu}
\affiliation{Kavli Institute for Cosmological Physics, University of
  Chicago, Chicago, IL 60637, USA}
\author{H.~Yang}
\affiliation{Department of Physics, Stanford University, Stanford, CA
  94305, USA}
\author{K.~W.~Yoon}
\affiliation{Department of Physics, Stanford University, Stanford, CA
  94305, USA}
 \author{E.~Young}
\affiliation{Kavli Institute for Particle Astrophysics and Cosmology,
  SLAC National Accelerator Laboratory, 2575 Sand Hill Rd, Menlo Park,
  CA 94025, USA}
\affiliation{Department of Physics, Stanford University, Stanford, CA
  94305, USA}
\author{C.~Yu}
\affiliation{Department of Physics, Stanford University, Stanford, CA
  94305, USA}
\author{L.~Zeng}
\affiliation{Harvard-Smithsonian Center for Astrophysics, 60 Garden
  Street MS 42, Cambridge, Massachusetts 02138, USA}
\author{C.~Zhang}
\affiliation{Department of Physics, California Institute of
  Technology, Pasadena, California 91125, USA}

%\collaboration{BICEP/\emph{Keck Array} collaboration}

\date{\today}

\begin{abstract}
We present a search for axion-like polarization oscillations in the cosmic microwave background~(CMB) with observations from the \emph{Keck Array}. A local axion field induces an all-sky, temporally sinusoidal rotation of CMB polarization. A CMB polarimeter can thus function as a direct-detection experiment for axion-like dark matter. We develop techniques to extract an oscillation signal. Many elements of the method are generic to CMB polarimetry experiments and can be adapted for other datasets. As a first demonstration, we process data from the 2012 observing season to set upper limits on the axion-photon coupling constant in the mass range $10^{-21}$-$10^{-18}~\mathrm{eV}$, which corresponds to oscillation periods on the order of hours to months. We find no statistically significant deviations from the background model. For periods larger than $24~\mathrm{hr}$ (mass~$m < 4.8 \times 10^{-20}~\mathrm{eV}$), the median 95\%-confidence upper limit is equivalent to a rotation amplitude of $0.68^\circ$, which constrains the axion-photon coupling constant to $g_{\phi\gamma} < \left ( 1.1 \times 10^{-11}~\mathrm{GeV}^{-1} \right ) m/\left (10^{-21}~\mathrm{eV} \right )$, if axion-like particles constitute all of the dark matter. The constraints can be improved substantially with data already collected by the BICEP series of experiments. 
Current and future CMB polarimetry experiments are expected to achieve sufficient sensitivity to rule out unexplored regions of the axion parameter space.
\end{abstract}

%\keywords{}

\maketitle

\section{Introduction \label{sec:intro}}

With many astrophysical and cosmological measurements establishing the existence of dark matter, an understanding of its particle properties is one of the main aspirations of modern physics~\cite{Bergstrom2000,Bertone2005}. A promising dark-matter candidate is the QCD (quantum chromodynamics) axion, which we define here to be the pseudo-Nambu-Goldstone degree of freedom associated with the Peccei-Quinn mechanism that has been proposed to solve the strong CP problem~\cite{PQ1977Jun,PQ1977Sep,Weinberg1978,Wilczek1978,Preskill1982,Abbott1982,Dine1982}. In this work, we consider the much larger class of \emph{axion-like particles} (sometimes abbreviated as ALPs), which are light, bosonic particles with couplings to the Standard Model (SM) that are similar to that of the QCD axion but with important differences. Whereas the QCD axion requires a specific coupling to the QCD field strength that generically gives rise to a relationship between the mass of the QCD axion and its coupling to the SM, albeit with some model dependence, the axion-like particles considered in this work lack this coupling to QCD. They are, therefore, not related to solutions of the strong CP problem and generally have no fixed relationship between their mass and coupling to the SM. Because of this, they occupy a much larger area in the mass-coupling parameter space. For simplicity, we will hereafter refer to axion-like particles as \emph{axions}.

Very light axions can have astrophysically large de-Broglie wavelengths, which have macroscopic consequences for the formation of structure. Such dark-matter candidates are sometimes referred to as \emph{fuzzy dark matter} (FDM)~\cite{Hui2017}. 

An important property of an axion field is that it creates an effective birefringence for opposite-helicity photons. Linear polarizations are, therefore, rotated, and the amount of rotation is proportional to the change in the axion field between the point of emission and the point of absorption~\cite{Carroll1990,Carroll1991,Harari1992,Carroll1998}. In particular, we emphasize that the rotation of polarization responds to the harmonic oscillations of the axion field that occur with a frequency~$m/(2\pi)$, where $m$~is the axion mass. In this paper, we consider axion masses in the range of $10^{-21}$-$10^{-18}~\mathrm{eV}$, which roughly corresponds to oscillation periods of hours to months.

Recently, Fedderke et al. proposed two axion observables accessible by current and future cosmic-microwave-background~(CMB) polarimetry experiments~\cite{Fedderke2019}. The first is an overall suppression of CMB polarization, which is referred to as the ``washout'' effect and can be constrained by measurements of the $TT$, $TE$ and $EE$ power spectra. The second is a time-varying global rotation of CMB polarization with angular frequency~$m$. The latter observable is called the ``AC oscillation'' and is the main focus of this work.

The washout effect is due to the axion-field evolution during the epoch of recombination. With this observable, Fedderke et al. used publicly available \emph{Planck} data to rule out regions of the axion parameter space, which we show below in \figref{fig:exclusion}. The washout is, ultimately, cosmic-variance limited, because it relies on the statistics of power spectra. The limits already set with the washout effect are within an order of magnitude of the cosmic-variance limit. 

The AC oscillation is a sinusoidal global rotation of CMB polarization with an angular frequency~$m$. Whereas the washout effect is sensitive to axion dark matter present during the epoch of recombination, the AC oscillation is sensitive to axion dark matter at the location of the experiment. The temporal change in CMB polarization is a direct probe of the oscillation of the \emph{local} axion field. The measurement of axion-like polarization oscillations in the CMB is a form of direct dark-matter detection. The expected coherence time is $\sim 2 \pi /(m v^2)$, where $v \sim 10^{-3}$~is the Galactic virial velocity. The associated coherence \emph{length} is $\sim 2 \pi/(m v)$. For oscillation periods shorter than $\sim 1~\mathrm{day}$, existing axion limits are stronger than what can be achieved with the current generation of CMB instruments, so we set a minimum oscillation period of~$2~\mathrm{hr}$ for our search.  For $2\pi/m = 2~\mathrm{hr}$, the coherence time is $\sim 200~\mathrm{yr}$, and the coherence length is $\sim 0.07~\mathrm{pc}$. We can, therefore, take the oscillation to be in phase for all CMB experiments. Furthermore, the signal should be in phase across all photon frequencies. The data from experiments operating at different times, locations and wavelengths can be combined to search for a coherent polarization oscillation. As the signal persists in time, there is no cosmic-variance limit. In the long term, therefore, the oscillation effect will likely be a more sensitive observable than the washout effect.

Denote by~$Q(\unitVec{n},t)$ and~$U(\unitVec{n},t)$ the Stokes parameters that are measured at sky coordinate~$\unitVec{n}$ at time~$t$. Denote by~$Q_0(\unitVec{n})$ and~$U_0(\unitVec{n})$ the Stokes parameters that would be measured if the axion were completely decoupled from photons, which is also the limit in which the CMB polarization field is static. The axion-photon coupling constant is~$g_{\phi \gamma}$. The amplitude of the axion field averaged over the CMB visibility function is denoted~$\avg{\phi_\ast}$, which we take to be isotropic. The amplitude of the local axion field today is denoted~$\phi_0$. We allow for an arbitrary phase~$\alpha$ in the oscillation. Then the observed Stokes parameters are related to the decoupled limit by~\cite{Fedderke2019}
\begin{widetext}
\eq{ Q(\unitVec{n},t) \pm i U(\unitVec{n},t) = J_0(g_{\phi \gamma} \avg{\phi_\ast}) \exp \brackets{ \pm i g_{\phi \gamma} \phi_0 \cos (m t + \alpha) } \parens{ Q_0(\unitVec{n}) \pm i U_0(\unitVec{n}) } . }
\end{widetext}
All of the combinations of the form~$g_{\phi \gamma} \phi_x$ for $x \in \{ \ast,0 \}$ are small and dimensionless. The Bessel function can be expanded as $J_0(x) \approx 1 - x^2/4$ and represents an overall suppression of CMB polarization, i.e., the washout effect described above. Expanding the complex exponential to first order and defining the time-averaged polarization fields~$\Qavg{}$ and~$\Uavg{}$, we have
\eq{ \parens{ \begin{array}{c} Q(\unitVec{n},t) \\ 
					U(\unitVec{n},t) \end{array} }  = 
					\parens{ \begin{array}{cc} 1 & -f(t) \\
										f(t) & 1 \end{array} }
					\parens{ \begin{array}{c} \Qavg{} \\
										\Uavg{} \end{array} } , \label{eq:QUmixingmatrix} }
where 
\eq{ f(t) \equiv g_{\phi \gamma} \phi_0 \cos(m t + \alpha) . \label{eq:f(t)functionalform} }
To linear order, the polarization oscillation causes a mixing of Stokes parameters. The parameter~$f(t)$ is small compared with unity, so the mixing matrix in \eqref{eq:QUmixingmatrix} can be viewed as a \emph{rotation} matrix expanded to leading order. The time-averaged fields~$\Qavg{}$ and~$\Uavg{}$ are rotated into each other by an angle~$f(t)$. This is equivalent to an on-sky rotation of the polarization pseudovectors by an angle~$f(t)/2$. In \eqref{eq:QUmixingmatrix}, the washout effect has been absorbed into the definition of the time-averaged polarizations fields, i.e., $\Qavg{} = J_0(g_{\phi \gamma} \avg{\phi_\ast}) Q_0(\unitVec{n})$ and similarly for Stokes~$U$. Since the time-averaged polarization fields are direct observables of CMB polarimetry experiments, we can use \eqref{eq:QUmixingmatrix} to search for the Stokes mixing~$f(t)$ without referring to the decoupled limit, i.e., to~$Q_0(\unitVec{n})$ and~$U_0(\unitVec{n})$.

Although relatively faint compared with many polarized astrophysical sources, the CMB provides a number of serendipitous advantages in the search for axion-like polarization oscillations. Current-generation CMB experiments have deployed thousands of photon-noise-limited detectors that scan CMB-dominated patches of sky repetitively for years. The steady increase in detector count in CMB experiments translates directly to an increase in the statistical weight of each instantaneous measurement. Since the signal is a coherent all-sky rotation of polarization angles, every optically active detector can contribute to the measurement. The detectors are observing the CMB for a substantial fraction of each year, which provides temporal sensitivity on timescales of hours to years. By repetitively scanning the same patch of sky, the time-averaged maps~$\Qavg{}$ and~$\Uavg{}$ can be well estimated and used as templates to search for time-variability as in \eqref{eq:QUmixingmatrix}. The oscillation signal is coherent both over the sky and over wavelength, so all CMB instruments can contribute independent of angular resolution and observing frequency. 

The CMB has a theoretical advantage in that the axion field at the point of emission is effectively zero~\cite{Fedderke2019}. The surface of last scattering represents an era much longer ($\mathcal{O}(10^4)~\mathrm{yr}$) than the axion oscillation periods under consideration in this work ($\mathcal{O}(1)~\mathrm{yr}$). Emissions from different redshifts occur with different axion field values that, taken together, average to approximately zero along all lines of sight.
An oscillation observed in the CMB today is, therefore, a direct measure of the \emph{local} axion field only, i.e., the field at the point of \emph{absorption}. 

Polarization oscillations may also be observed in astrophysical sources such as pulsars~\cite{Caputo2019,Liu2019}, the jets of active galaxies~\cite{Ivanov2018}, protoplanetary disks~\cite{Fujita2018} and strong gravitational lens systems~\cite{Basu2020}. A complication in setting constraints with polarized astrophysical sources is the uncertainty in both the amplitude and the phase of the axion field at the point of \emph{emission}, and these amplitudes and phases are, in general, different for different sources, which may introduce a large number of free parameters. A CMB-based search is constrained to have the same amplitude and phase across the entire sky, at every wavelength and at every observing site.

One of the main challenges in many CMB polarimetry experiments is contamination from Galactic foregrounds. These foregrounds are less problematic in an axion-oscillation search for two reasons. The first is that the oscillations affect \emph{all} CMB polarization, i.e., both $E$- and $B$-modes. While the Galactic foregrounds dominate at, e.g., $150~\mathrm{GHz}$ in $B$-modes even in the cleanest patches of sky, they are subdominant in $E$-modes. In the BICEP observation patch, the polarization power in foregrounds is $\sim 10\%$ as strong as the CMB power. By considering the much brighter $E$-modes, we make the foreground contamination relatively weaker. The second reason is that the foregrounds do not present an all-sky coherent polarization oscillation. Some polarized signals from the Galaxy may be emitted from regions with a substantially different axion field value, and these signals would oscillate. The amplitude and phase of these oscillations, however, would depend on the axion field value at the point of emission, and these axion field values would not be coherent across the entire observing region. By constraining our search to polarization oscillations that are both global and coherent, we suppress contamination from Galactic foregrounds. For these reasons, we consider foreground contamination to be a minor concern.

Polarization oscillations are sensitive to the product~$g_{\phi\gamma} \phi_0$ (\eqref{eq:f(t)functionalform}) and, therefore, depend on both the axion-photon coupling constant~$g_{\phi\gamma}$ and the axion mass~$m$. The latter dependence comes from the axion field strength
\spliteq{ \phi_0 & = \parens{2.1 \times 10^9~\mathrm{GeV}} \parens{ \frac{m}{10^{-21}~\mathrm{eV}} }^{-1} \\
& \quad\quad \times \parens{ \frac{\kappa \rho_0}{0.3~\mathrm{GeV}/\mathrm{cm}^3}}^{1/2} , \label{eq:phi0} }
where $\rho_0$~is the local density of dark matter and $\kappa$~is the fraction of dark matter composed of axion-like particles~\cite{Fedderke2019}. The $m$-dependence of~$\phi_0$ implies that oscillation-derived limits on the coupling constant will roughly follow $g_{\phi\gamma} \propto m$. The coupling constant can be probed approximately independently of mass in a number of ways, e.g., by conversion of solar axions to x-rays in strong laboratory magnetic fields~\cite{CAST2017}, by conversion of supernova-produced axions to gamma rays in Galactic magnetic fields~\cite{Payez2015}, from the x-ray transparency of the intracluster medium~\cite{Reynolds2019} (though we note that this bound has been challenged~\cite{Libanov2019}) and by conversion of axions produced in Wolf-Rayet stars to x-rays in Galactic magnetic fields~\cite{Dessert2020}. At the same time, the axion mass can be constrained approximately independently of the coupling constant through considerations of small-scale structure, e.g., in the Lyman-$\alpha$ forest~\cite{Irsic2017}, in the population of Milky Way satellite galaxies~\cite{Nadler2019} and with the subhalo mass function~\cite{Schutz2020}. These investigations have set similar bounds in the range of $m \gtrsim 2 \times 10^{-21}~\mathrm{eV}$, which suggests a maximum oscillation period of~$\sim 20~\mathrm{d}$. Ultimately, the axion parameter space will be constrained by a variety of probes, each subject to a different set of systematic uncertainties.

\begin{figure*}
	\includegraphics[width = \textwidth]{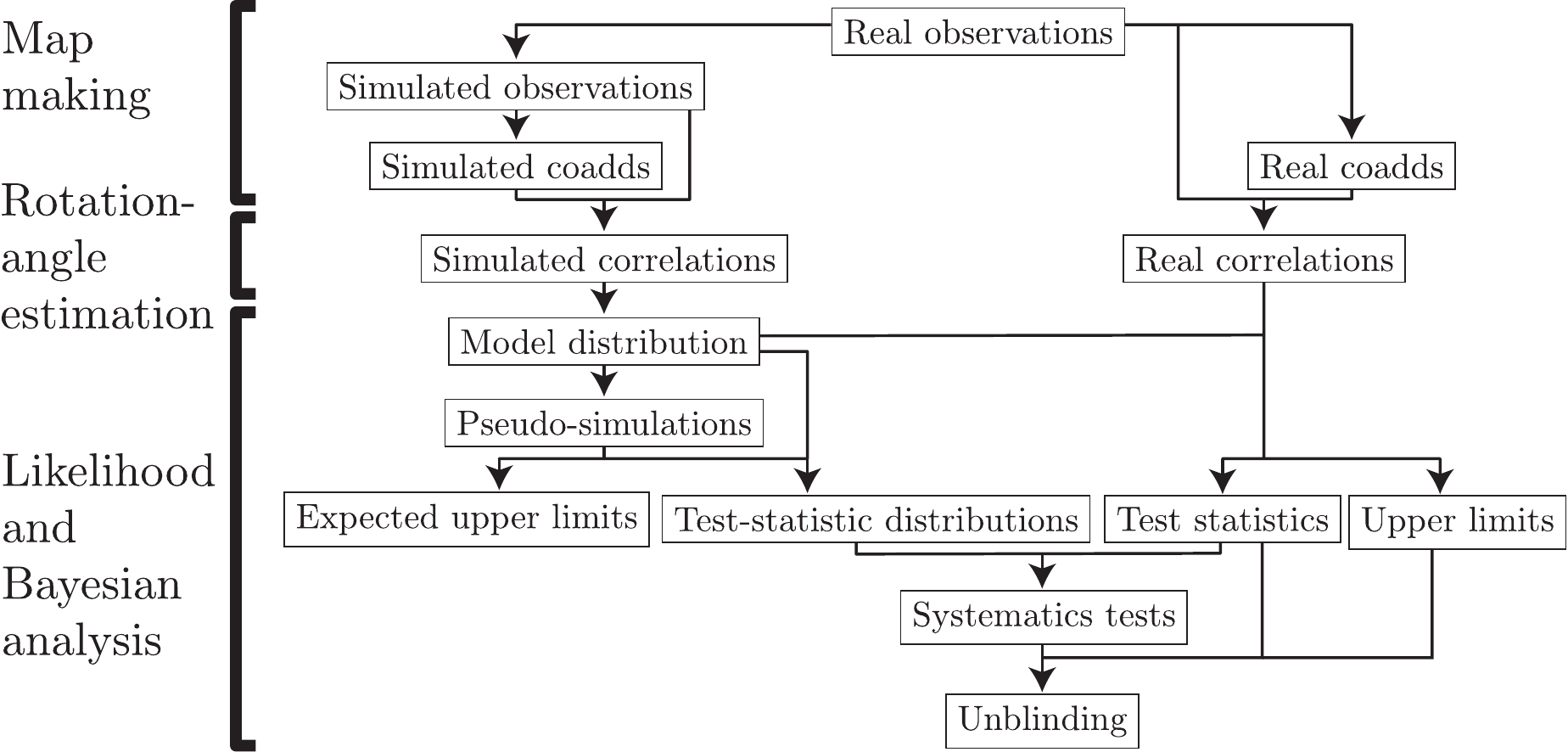}
	\caption{Flow diagram of the axion-oscillation analysis pipeline. The three main components are indicated on the left, and more detail is presented in the rest of the figure. The map-making step is similar to that of a standard CMB analysis, though the reobservation must now include a component that represents polarization oscillations (Secs.~\ref{sec:multipleComponents}, \ref{sec:observables} and~\ref{sec:sims}). The coadds are estimates of the time-averaged Stokes parameters~$\Qavg{}$ and~$\Uavg{}$. The global rotation angle~$f(t)/2$ is obtained through the \emph{correlation method} of \secref{sec:corrMethod} with some of the computational speed-ups of \secref{sec:sims}. The likelihood and Bayesian analysis are described in \secref{sec:likelihood}. The ensemble of simulations is used to construct a model distribution, which we can resample to form large numbers of Gaussian pseudo-simulations (\secref{sec:pseudosims}). The model distribution also implies a likelihood function (\secref{sec:likelihoodFunction}) that is used to set Bayesian upper limits (\secref{sec:BayesianUpperLimits}). At the same time, we estimate test statistics to check consistency with the background model (\secref{sec:bkgConsistencyTest}) and to test for spurious systematic signals (\secref{sec:systematics}). Based on the results of the systematics checks, which we call ``jackknife tests'', we unblind the real non-jackknife data (\secref{sec:unblinding}). \label{fig:analysis_pipeline}}
\end{figure*}

The paper is organized as follows.
In \secref{sec:instrument}, we provide an overview of the BICEP program including details of the instruments and the integrated dataset. The next several sections outline our analysis method to search for axion-like polarization oscillations. Some of the analysis choices are specific to the \emph{Keck Array}, and some expectations are stated on the basis of experience with the BICEP dataset. Many elements of the method, however, can be adapted for other CMB polarimetry experiments. An overview of the analysis structure is shown in \figref{fig:analysis_pipeline}.
There are three main components of the analysis. 
The first step is to make maps, both real and simulated, which we do in the same way as for a standard CMB analysis~\cite{BK-I}. These maps are our best estimates for the time-averaged Stokes parameters~$\Qavg{}$ and~$\Uavg{}$. The main departure from standard CMB map making is the reobservation of a simulation component that represents a global polarization rotation. We call this component the ``rotated  CMB map'', which we will often abbreviate as~$\mathrm{rCMB}$, and its computational utility is described in \secref{sec:sims}. 
In \secref{sec:rotEst}, we describe our method for extracting estimates of the global polarization rotation angle~$f(t)/2$ as a function of time. For this purpose, we introduce the correlation method of \secref{sec:corrMethod}. 
From our ensemble of simulations, we construct a model distribution and an associated likelihood function. The statistical analysis is presented in \secref{sec:likelihood}. We impose conservative prior distributions on the oscillation parameters and extract Bayesian upper limits on the oscillation amplitude. Additionally, we search for systematics in our data with several ``jackknife tests'' described in \secref{sec:systematics}. Having presented our analysis methods, we then use the 2012 observing season of the \emph{Keck Array} for a first demonstration of the techniques, and the results are given in \secref{sec:results}. We close in \secref{sec:conclusions} with some expectations for future results from BICEP and from other current and planned CMB experiments.

%%%
\section{Instrument overview \label{sec:instrument}}

The \emph{Keck Array} observed from the South Pole from 2012 to 2019 and consisted of a single mount with five microwave receivers, each similar to the precursor BICEP2~\cite{BK-II}. Each receiver is an independent refracting telescope with cryogenic lenses and an aperture diameter of $25~\mathrm{cm}$. The entire mount can be rotated to change the boresight orientation or \emph{deck angle} of all five receivers. Each focal plane consists of 512 dual-polarized slot-dipole antenna arrays coupled to transition-edge-sensor (TES) bolometers~\cite{BKSPIDER2015}. A SQUID-based time-division multiplexing system is used to read out the TESs~\cite{deKorte2003}. Each pixel is coupled to two TESs, one for each linear polarization, and the difference in signals is a measure of the on-sky polarization.

The main observing region occupies $\sim 1\%$ of the sky centered on $\mathrm{RA}~0\mathrm{h}$, $\mathrm{Dec.}~-57.5^\circ$. For the 2012 and 2013 observing seasons, all five receivers observed at~$150~\mathrm{GHz}$. In subsequent seasons, some receivers were switched to other observing frequencies. In the results below, however, we consider data from only the 2012 season.

The main science goal of the BICEP program is the search for $B$-mode polarization from primordial gravitational waves~\cite{Kamionkowski1997,Seljak1997}. In combination with \emph{Planck}~\cite{Planck2015I} and WMAP~\cite{WMAP9}, the BICEP2/\emph{Keck Array} experiments have detected $B$-modes from gravitational lensing~\cite{Zaldarriaga1998} at $8.8\sigma$ significance and constrained the tensor-to-scalar ratio to $r_{0.05} < 0.06$ using data through the 2015 season~\cite{BK-X}. The polarization map depth achieved by this dataset at $150~\mathrm{GHz}$ is $2.9~\mu\mathrm{K}_\mathrm{CMB}~\mathrm{arcmin}$. The results presented in \secref{sec:results} use only data from a single season, but we intend to extend this axion-oscillation analysis to the full BICEP dataset in the future and take advantage of the full sensitivity of these polarization maps. The complete dataset also includes maps at~$95$ and $220~\mathrm{GHz}$, which have achieved polarization map depths of~$5.2$ and $26~\mu\mathrm{K}_\mathrm{CMB}~\mathrm{arcmin}$, respectively. All frequencies with CMB sensitivity can be used for the axion-oscillation search.

Although designed for other purposes, the BICEP instruments, scan strategy and data processing are compatible with an axion-oscillation search. The methods and results presented below rely on data taken from standard observations targeted at CMB $B$-mode polarization. No change to the scan strategy or low-level data processing is necessary for an axion-oscillation analysis.

%%%
\section{Global polarization rotation estimation \label{sec:rotEst} }

The BICEP experiment does not measure Stokes parameters instantaneously but instead measures \emph{pair differences}, i.e., the difference in power between two orthogonally polarized detectors. The average Stokes parameters are constructed only after repeated observations have been made with multiple detector orientations. From a single pair-difference measurement, it is not possible to construct Stokes parameters, but we will argue that it is not necessary for estimating~$f(t)$.

%%%
\subsection{Pair difference}

We parameterize the polarization orientation of detector pair~$i$ by the angle~$\psi_i(\unitVec{n},t)$.\footnote{There is some redundancy in the expression for the polarization orientation~$\psi_i(\unitVec{n},t)$, since a knowledge of the detector pointing as a function of time immediately implies a value for~$\unitVec{n}$ given~$t$ and~$i$. We will, however, bin observations in time. In each time bin, labeled by some mean time~$\tau$, a detector pair produces a map that covers many sky coordinates~$\unitVec{n}$, so it is useful to keep track of both the mean time~$\tau$ and the sky coordinate~$\unitVec{n}$.}  
The pair difference is then related to the Stokes parameters by
\eq{ \Dit{} = Q(\unitVec{n},t) \cit{} + U(\unitVec{n},t) \sit{} , \label{eq:DrelatedtoQU} }
where we introduce the shorthand
\aligneq{ \cit{} & \equiv \cos[2 \psi_i(\unitVec{n},t)] , & \sit{} & \equiv \sin[2 \psi_i(\unitVec{n},t)] . \label{eq:citdef sitdef} }
In the limit that the polarization field is dominated by the CMB, we can decompose the pair difference into two components: a static component~$\Distatict$ that depends on the average Stokes parameters and an oscillating component~$\Diosct$ that is induced by the local axion field. Combining Eqs.~\ref{eq:QUmixingmatrix} and~\ref{eq:DrelatedtoQU}, we write
\eq{ \Dit{} = \Distatict + \Diosct , \label{eq:Dit decomposed} }
where 
\eq{ \Distatict = \Qavg{} \cit{} + \Uavg{} \sit{} \label{eq:Distatict} }
and
\eq{ \Diosct = f(t) \parens{ \Qavg{} \sit{} - \Uavg{} \cit{} } . \label{eq:Diosct} }
The ``static'' component depends on time only through the time-dependence of the polarization orientation~$\psi_i(\unitVec{n},t)$. We call it ``static'', because the underlying Stokes parameters are static. The average Stokes parameters~$\Qavg{}$ and~$\Uavg{}$ are standard data products of the BICEP experiment and can be considered, at least approximately, known quantities. The trigonometric factors~$\cit{}$ and~$\sit{}$ depend on detector pointing and orientation and are also known. The full pair difference~$\Dit{}$ is measured, so the only unknown quantity is~$f(t)$.

We define the \emph{rotated map}
\eq{ \rit{} \equiv \Qavg{} \sit{} - \Uavg{} \cit{} \label{eq:ritDef} }
and note
\eq{ \Diosct = f(t) \rit{} . \label{eq:Dosc = f r} }
The rotated map~$\rit{}$ is a template that can be used to search for an oscillating component of the measured pair difference. We correlate the rotated map~$\rit{}$ with the pair difference~$\Dit{}$ to estimate~$f(t)$.

Comparing Eqs~\ref{eq:DrelatedtoQU} and~\ref{eq:ritDef}, we see that the rotated map~$\rit{}$ is the pair difference that would be measured from Stokes parameters that are orthogonal (in $Q$/$U$-space) to the time-averaged values~$\Qavg{}$ and~$\Uavg{}$.
This orthogonal component contributes~$\Diosct$ (\eqref{eq:Dosc = f r}) to the instantaneous pair difference~$\Dit{}$  when there is a global rotation of the time-averaged polarization field by an angle~$f(t)/2$.

%%%
\subsection{Multiple components \label{sec:multipleComponents} }

The above discussion assumed the polarization field is dominated by the CMB. We now take a more realistic approach and include noise and foregrounds.\footnote{In \secref{sec:intro}, we argued that foreground contamination is expected to be negligible. We, therefore, make the simplifying approximation that the polarized foregrounds are not themselves subject to axion-induced polarization oscillations.} Then the time-dependent Stokes fields can be expressed as
\spliteq{ Q(\unitVec{n},t)  & = \QavgCMB + Q^{(\mathrm{fg})}(\unitVec{n}) + Q^{(\mathrm{N})}(\unitVec{n},t) \\
& \quad\quad + Q^{(\mathrm{osc})}(\unitVec{n},t) . }
and similarly for~$U(\unitVec{n},t)$, where the terms on the right-hand side represent, respectively, time-averaged CMB, static foregrounds, time-varying noise and time-varying polarization oscillations. The pair difference is linearly related to these components and can also be written as a sum of the four contributions:
\spliteq{ \Dit{} & = \Dicmbt + \Difgt + \Dinoiset \\
& \quad\quad + \Diosct . } 
The instantaneous pair difference~$\Dit{}$ is dominated by the noise term~$\Dinoiset$, so we approximate the variance in our measurement of~$\Dit{}$ to be entirely due to noise. The next-largest contribution is from the time-averaged CMB and is mainly in the form of $E$-modes. Next, we have foregrounds. For $100$-$300~\mathrm{GHz}$, the dominant foreground is Galactic dust, which is suppressed relative to the CMB $E$-modes by approximately an order of magnitude in the BICEP observation region. Finally, we have the oscillation signal~$\Diosct$. It will be useful to define the quantities
\aligneq{ Q^{(\mathrm{rCMB})}(\unitVec{n}) & \equiv -\UavgCMB , \\ U^{(\mathrm{rCMB})}(\unitVec{n}) & \equiv \QavgCMB }
as well as the pair difference formed from these ``rotated'' Stokes parameters, which is given by the usual formula
\eq{ \Dircmbt \equiv Q^{(\mathrm{rCMB})}(\unitVec{n}) \cit{} + U^{(\mathrm{rCMB})}(\unitVec{n}) \sit{} . }
Then the oscillating component of the pair difference is 
\eq{ \Diosct = f(t) \Dircmbt . \label{eq:DoscRelatedToDrcmb} }
This is a useful recasting, since the quantity~$\Dircmbt$ depends only on the time-averaged CMB polarization field and the detector pointing. This will become especially convenient in our discussion of efficient simulation schemes. The quantity~$\Dircmbt$ is on the order of the static CMB, and the smallness of the oscillating component~$\Diosct$ is, therefore, made manifest by the factor~$f(t)$. Since an oscillation amplitude of~$0.1^\circ \approx 2 \times 10^{-3}$ has already been ruled out with \emph{Planck} data~\cite{Fedderke2019}, the oscillating component of the pair difference is suppressed relative to the CMB $E$-modes by at least $\sim 10^{2}$.

Our goal is to extract the component~$\Diosct$ from the total pair difference~$\Dit{}$. We argued above that the oscillating component of the pair difference is proportional to~$\rit{}$ in the limit that the time-averaged maps are CMB-dominated, which is a good approximation at the map depths achieved by the BICEP/\emph{Keck Array} experiments. The rotated map is constructed from known and measured quantities. To extract the oscillating component, then, we correlate the instantaneous pair difference with the rotated map, i.e., we correlate~$\Dit{}$ with~$\rit{}$.

%%%
\subsection{Time binning \label{sec:timeBinning}}

The fundamental unit of observation in the BICEP/\emph{Keck Array} experiments is the \emph{scanset}, which consists of approximately 45~minutes of constant-elevation scanning. It is, therefore, convenient to bin observations by scanset. This makes the analysis insensitive to oscillation periods shorter than roughly $1~\mathrm{hr}$, but the constraints on the axion-photon coupling from measurements of SN1987A already rule out an observable polarization oscillation at these timescales given current-generation sensitivities~\cite{Payez2015}.\footnote{As discussed explicitly in \secref{sec:ULresults}, the oscillation analysis sets limits that scale as $g_{\phi\gamma} \propto m$, so sensitivity to the coupling constant~$g_{\phi\gamma}$ is worse at higher masses (shorter oscillation periods).} Consequently, there is little motivation to extend the analysis to these short periods.

In each time bin, we construct a pair-difference map for each detector, which we call a \emph{pairmap}. We take the variable~$\tau$ to label the mean time of a scanset, and we take the sky coordinate~$\unitVec{n}$ to be discretized according to the BICEP map pixelization~\cite{BK-I}. Then denote by~$\Ditau{}$ the pairmap constructed for detector~$i$ during the scanset that occured at mean time~$\tau$. This map will cover only a fraction of the full observation patch. The map pixels that are covered may have been visited multiple times over the course of the observation, and $\Ditau{}$~represents a weighted average of these repeated measurements. We also include in the definition of~$\Ditau{}$ any timestream filtering, so $\Ditau{}$~is the quantity that is coadded over all detectors and observations in a standard CMB analysis to form the final map.%
%%%
\footnote{The pair difference is not itself coadded. When a map pixel~$\unitVec{n}$ has been visited more than once with different polarization orientations~$\psi_i(\unitVec{n},t)$, it is possible to extract Stokes~$Q$ and~$U$ from the pair-difference measurements. The quantities that are actually used for the coadd are~$\cit{} \Dit{}$ and~$\sit{} \Dit{}$. With multiple visits at different orientations, these quantities can be inverted to recover~$Q(\unitVec{n})$ and~$U(\unitVec{n})$.} 

If the scan strategy involves multiple visits to the same map pixels, then the polarization angle~$\psi_i(\unitVec{n},t)$ may be different each time due to sky or, in principle, instrument rotation. If the pair differences from all of the visits are then averaged, the trigonometric factors~$\cit{}$ and~$\sit{}$ will beat against each other and wash out the polarization signal. An advantage of observing from the South Pole is that a constant-elevation scan is also approximately a constant-declination scan, so repeated visits to the same map pixel have nearly the same~$\psi_i(\unitVec{n},t)$, as long as no boresight rotation has been performed. Denote the weighted average of the trigonometric factors by~$\citau{}$ and~$\sitau{}$. The weights are the same as those used to form the scanset map~$\Ditau{}$, though there is no filtering.

For oscillation periods of the same order as the time binning, there will be a suppression of the axion signal. For a time bin of length~$\Delta t$, the average Stokes mixing angle is
\eq{ \ftau = f(\tau) \operatorname{sinc} \parens{ \frac{ m \Delta t}{2} } , \label{eq:avgfsinc} }
where the notation~$\ftau$ indicates the average value of~$f(t)$ in the scanset of mean time~$\tau$. The shortest oscillation period we will consider is~$2~\mathrm{hr}$, for which a 45-minute observation produces a suppression of~$22\%$. This is arguably small but should be accounted for in referring our measurements to an oscillation amplitude, since we would otherwise claim greater sensitivity at these short periods than is justified.

%%%
\subsection{Observables \label{sec:observables}}

In \secref{sec:multipleComponents}, we outlined a method of correlating the time-averaged CMB maps~$\QavgCMB$ and~$\UavgCMB$ with the instantaneous pair difference~$\Dit{}$ as a means of extracting the polarization rotation angle. We do not have direct access to the true time-averaged CMB maps. Instead, we construct coadded maps, which are non-trivially impacted by filtering operations. As these coadded maps are dominated by CMB polarization, we will use them as templates. We use the notation~$\Qcoadd{}$ to denote the measured coadd for Stokes~$Q$ and similarly for~$U$. These coadds include all of the biases introduced by weighting and filtering observations but are the best available approximations to the time-averaged sky. While the true time-averaged values of noise and oscillations are zero, our coadds will, in general, contain non-zero contributions from these components. Our coadd for Stokes~$Q$ is, then,
\eq{ \Qcoadd{} = \QcoaddCMB + \Qcoaddfg + \Qcoaddnoise + \Qcoaddosc }
and similarly for Stokes~$U$. Our approximation for the rotated map~$\rit{}$ is, then,
\eq{ \ritau{} \equiv \Qcoadd{} \sitau{} - \Ucoadd{} \citau{} , \label{eq:ritauDef} }
where $\tau$~is the nearest mean sub-observation time to~$t$.

%%%
\subsection{Correlation method \label{sec:corrMethod}}

We define the correlation
\eq{ \rho(\tau) \equiv \frac{1}{W(\tau)} \sum_{i,\unitVec{n}} \ritau{} \Ditau{} \witau{} \vitau{} , \label{eq:rhoDef} }
where
\spliteq{ \witau{} & \equiv \frac{1}{\var{\brackets{\Ditau{}}}} , \\
 	\vitau{} & \equiv \frac{1}{\var{\brackets{\ritau{}}}} , \\
  	W(\tau) &  \equiv \sum_{i,\unitVec{n}} \witau{} \vitau{} . \label{eq:weightDefinitions} }
The quantity~$\rho(\tau)$ is a weighted correlation of the measured pair difference~$\Ditau{}$ with our approximation to the rotated map~$\ritau{}$. The weights are the inverse variances of the two maps. In general, the variance in the per-scanset pair difference will be relatively large, being dominated by atmospheric fluctuations at the time of observation. This variance~$\witau{}$ is a direct observable, which we use to downweight noisy pair-difference measurements in our standard map-making pipeline. At the same time, we use~$\vitau{}$ to downweight regions of the map, for which we have relatively poor estimates of the CMB polarization field. The efficacy of the analysis depends on having time-averaged maps that are dominated by CMB polarization. In general, it will be the edges of the coadded maps that show significant residual noise, since these map pixels are visited much less frequently than those at the center. The particular linear combination of time-averaged Stokes parameters with which we are correlating is~$\ritau{}$, so we downweight by its variance
\spliteq{ \var\brackets{\ritau{}} & = \var\brackets{\Qcoadd{}} \sitau{2} + \var\brackets{\Ucoadd{}} \citau{2} \\
	& \quad\quad - 2 \cov{\Qcoadd{}}{\Ucoadd{}} \citau{} \sitau{} . }
All of the variances and covariances on the right-hand side are standard data products that are produced alongside the coadded maps. The quantity~$W(\tau)$ is nothing more than a normalization. These weights can be directly estimated from the data and can be considered known quantities.

An inverse-variance weighting may be suboptimal for the rotated map~$\ritau{}$, because the $E$-modes in our coadded maps~$\Qcoadd{}$ and~$\Ucoadd{}$ tend to be stronger than the noise. In the signal-dominated limit, it is better to use all of the available modes. A possible improvement to the analysis is to weight by $1/(1+N/S)$, where $S/N$~is a figure of merit for the $E$-mode signal-to-noise ratio. For noisy pixels, this is essentially an inverse-variance weighting. For large signal-to-noise ratios, however, we achieve nearly equal weighting. The prescription amounts to an approximate Wiener filter.

The weighting in \eqref{eq:rhoDef} does not include covariances between detectors~$i$ nor between map pixels~$\unitVec{n}$. One consequence is that the weight~$W(\tau)$ does not exactly predict the true inverse variance of~$\rho(\tau)$. It is computationally simpler, however, to calibrate the variances through Monte Carlo simulations as described below in \secref{sec:modelDist}. Omitting covariances in the definition of~$\rho(\tau)$ (\eqref{eq:rhoDef}) may degrade the sensitivity of the analysis, but it does not bias the results. With some simplifying approximations, it may be computationally practical to estimate such covariances, and this is a possible avenue for improvement in a future iteration of an axion-oscillation search.

\begin{figure}
	\includegraphics[width = 0.5\textwidth]{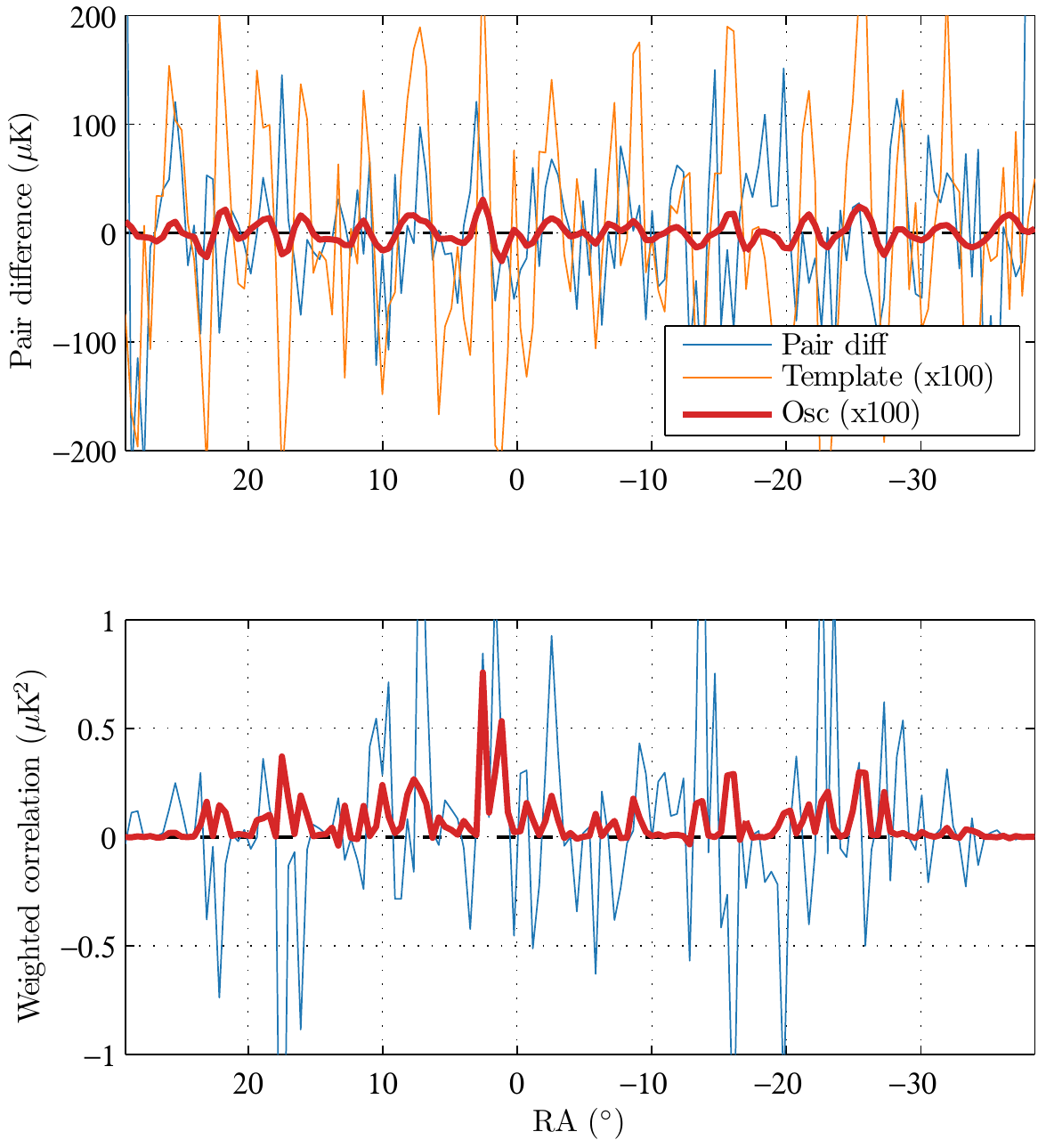}
	\caption{An illustration of the correlation method introduced in \secref{sec:corrMethod} for a simulation of a single detector and a single observation. (\emph{Top}) The pair difference~$\Ditau{}$ (blue) from a constant-elevation scan is plotted against right ascension (RA). The pair difference is dominated by atmospheric loading and fluctuates strongly. In this simulation, a global rotation of~$3^\circ$ has been imposed. The oscillating component~$\Diosctau$ (red) is hidden underneath the atmospheric fluctuations. The rotated map~$\ritau{}$ (orange) is used as a template to pick out global polarization rotations. (\emph{Bottom}) The correlation $\rho(\tau)$ (Eq.~\ref{eq:rhoDef} summed over detector~$i$ only) between the template and the full pair difference (blue) has a large variance but mean zero. The correlation with the oscillating component (red) is biased positive. \label{fig:rho_example}}
\end{figure}

We can model~$\Ditau{}$ as a linear combination of constituent components, i.e., 
\spliteq{ \Ditau{} & = \Dicmbtau + \Difgtau + \Dinoisetau \\ 
		& \quad\quad + \Diosctau , \label{eq:DitauDecomposed} }
where $\Ditau{(s)}$~is the weighted, filtered and binned scanset map formed from~$\Dit{(s)}$ for any component~$s$. We have direct access to these constituent components only in simulation. In the limit that the coadded maps are good representations of the true time-averaged polarization field and the pair-difference measurements are good probes of the instantaneous polarization field, we have, from \eqref{eq:Dosc = f r}, 
\eq{ \Diosctau \approx f(\tau) \ritau{} , }
when the oscillation period is sufficiently long to treat~$f(t)$ as constant over the scanset of mean time~$\tau$. We will discuss the case of faster oscillations in \secref{sec:sigTF}.
The correlation~$\rho(\tau)$ defined in \eqref{eq:rhoDef} picks out this component by correlating $\Ditau{}$ with~$\ritau{}$. On average, the correlation with the non-oscillating components of~$\Ditau{}$ vanishes approximately. Exact orthogonality is not necessary, though it improves the efficacy of the correlation method. An illustrative example of the correlation method is given in \figref{fig:rho_example}, where we show how the template~$\ritau{}$ picks out the oscillating component~$\Diosctau$ from the full pair difference~$\Ditau{}$.
The example shows only the results for a single detector~$i$ from a single scanset~$\tau$, and we see that the noise variance dominates. These noise fluctuations, however, average down when we combine many detectors over many scansets, while the correlation with the oscillating component, because it is biased positive, does not. In this way, we increase the signal-to-noise ratio as the dataset grows.

We define the normalizing constant
\eq{ R(\tau) \equiv \frac{1}{W(\tau)} \sum_{i,\unitVec{n}} \ritau{2} \witau{} \vitau{} , \label{eq:Rdef} }
in terms of which we form the estimator
\eq{ \hat{f}(\tau) \equiv \frac{\rho(\tau)}{R(\tau)} . \label{eq:fhatDef} }
Due to residual noise in the coadded maps, filtering, weighting, binning and imperfect orthogonality of the pair-difference components of \eqref{eq:DitauDecomposed}, the estimator must be calibrated through simulation. When these effects are negligible, however, $\hat{f}(\tau)$~is an unbiased estimator of~$f(\tau)$. 

%%%
\subsubsection{Optimality}

In the limit of Gaussian noise, negligible covariances and CMB-dominated template maps~$\Qcoadd{}$ and~$\Ucoadd{}$, it can be shown that $\hat{f}(\tau)$~(\eqref{eq:fhatDef}) is the maximum-likelihood estimator. We expect the assumption of Gaussian noise to be a good approximation. We expect detector-detector and pixel-pixel covariances to be small but potentially worth including in future iterations of the analysis. For one season of data from the \emph{Keck Array}, however, the noise power in the coadded polarization maps is about half as strong as the CMB power. When we extend the analysis to include more seasons of BICEP data, the assumption of CMB domination will become significantly better. For the preliminary results presented in this work, we accept the sensitivity hit from having an estimator that maximizes the likelihood only approximately. There may be some gains from accounting for covariances between detectors and between map pixels. As mentioned above, we also expect an improvement in sensitivity by Wiener filtering instead of inverse-variance weighting the rotated map~$\ritau{}$.

By comparing with results from the $EB$~nulling procedure of our standard CMB analysis pipeline~\cite{Kaufman2014}, we can roughly check the optimality of the correlation method as defined above. To absolutely calibrate the polarization angle of our receivers, we perform a global rotation to minimize the $EB$ and $TB$~cross spectra, which are expected to vanish in the CMB in the absence of cosmic birefringence. Finding the rotation angle~$\alpha_{EB}$ that minimizes~$EB$ only is similar to a search for axion-like polarization oscillations with $m = 0$, i.e., a temporally constant offset in~$\hat{f}(\tau)$.\footnote{A possible scheme for an axion-oscillation search is to estimate $EB$~rotation angles from small subsets of observations and look for time variability in~$\alpha_{EB}$. We considered but avoided this approach, because the incomplete map coverage of each scanset makes it awkward to form non-local quantities like the $E$- and $B$-modes. Additionally, we considered the $E$/$B$-decomposition and the associated cross-spectra to be unnecessary computational expenses for the purposes of detecting a time-varying global polarization rotation.} The uncertainty in~$\alpha_{EB}$ is a measure of the sensitivity of the $EB$~nulling procedure to an $m = 0$ oscillation. For the 2012 observing season of the \emph{Keck Array}, we find $\Delta\alpha_{EB} = 0.21^\circ$. In \secref{sec:bkgConsistencyTest} below, we outline a procedure for finding the best-fit oscillation amplitude and phase for each angular frequency~$m$. We can estimate an uncertainty by calculating the standard deviation of the constant offset for an ensemble of simulations. The result for $m = 0$ is~$0.27^\circ$. A comparison with the $EB$~result suggests the correlation method may be $\sim 30\%$ suboptimal, if none of the above improvements are implemented. The constant offset can be considered one oscillation mode, and each non-zero frequency $m \not = 0$ represents a single oscillation mode as well. We expect roughly similar sensitivity to each mode. We compare to the RMS rotation angle~$\theta_m$ for modes that pass through many oscillation periods in our dataset. If the rotation amplitude is~$\hat{A}_m/2$, then the RMS rotation angle (averaged over time) is~$\theta_m = \hat{A}_m/2^{3/2}$. For oscillation periods between~1 and 30~days, the real data show an RMS (averaged over~$m$) of $\Delta\theta = 0.28^\circ$, which again suggests a $\sim 30\%$ degradation relative to the $EB$~result. These comparisons are meant to give some indication of the possible margin for improvement. A full simulation-based study will be necessary to evaluate the true gains in sensitivity from, e.g., the changes suggested above, and we intend to report on this exploration in future publications.

%%%
\subsubsection{Correlation matrices \label{sec:corrMatrices}}

In simulation, it is useful to keep track of the independent components contributing to both the per-scanset pair difference~$\Ditau{}$ and to the rotated map~$\ritau{}$. We decomposed~$\Ditau{}$ in \eqref{eq:DitauDecomposed} into four components: time-averaged CMB, foregrounds, time-varying noise and time-varying polarization oscillations. As the map-making process is linear in the map components, we can decompose the rotated map, which is built from coadded maps, in the same way. We make the approximation that the contribution of polarization oscillations to the coadd is negligible, since we expect it to be small and to average down over all the observations included in the coadd. Then the rotated map can be decomposed as
\eq{ \ritau{} = \ricmbtau + \rifgtau + \rinoisetau , \label{eq:rOnly3components} }
where 
\eq{ \ritau{(s)} = \Qcoadd{(s)} \sitau{} - \Ucoadd{(s)} \citau{} . }
We approximate the weights to be dominated by the noise component, i.e., 
\aligneq{ \witau{} & \approx \frac{1}{\var\brackets{\Dinoisetau}} , \\ \vitau{} & \approx \frac{1}{\var{\brackets{\rinoisetau}}} . }
In particular, we note that these weights are independent of the non-noise components.

With all of these constituent components, the correlation~$\rho(\tau)$ can be written as a sum of cross terms, i.e., 
\eq{ \rho(\tau) = \sum_{s_r,s_D} \rhosrsD , }
where
\eq{ \rhosrsD \equiv \frac{\sum\limits_{i,\unitVec{n}} \ritau{(s_r)} \Ditau{(s_D)} \witau{} \vitau{} }{W(\tau)} . \label{eq:rhosrsD} }
On a per-observation basis, we expect $\rho(\tau)$~to be dominated by~$\rhocmbnoise$, since the rotated map is dominated by~$\ricmbtau$ and the pair difference by~$\Dinoisetau$. On average, we expect $\rhocmbnoise$~to vanish, but it has the largest variance of all the elements in the $\rho$-matrix. The oscillation signal is in~$\rhocmbosc$. The other elements make relatively minor contributions.

We also decompose the normalization constant as
\eq{ R(\tau) = \sum_{s_{r_1},s_{r_2}} \Rsrsr , }
where
\eq{ \Rsrsr \equiv \frac{\sum\limits_{i,\unitVec{n}} \ritau{(s_{r_1})} \ritau{(s_{r_2})} \witau{} \vitau{} }{W(\tau)} . \label{eq:Rsrsr} }
We can consider~$\Rsrsr$ to be a symmetric matrix at each observation time~$\tau$. The matrix is dominated by the diagonal terms, which are essentially autocorrelations of the constituent components of the rotated map~$\ritau{}$. As the CMB $E$-modes dominate the coadded maps, we expect the largest contribution to be from~$\Rcmbcmb$. The next largest contributions will be from the residual map noise~$\Rnoisenoise$ and the foregrounds~$\Rfgfg$.

The correlation matrices are dominated by a relatively small minority of elements. Roughly, we expect
\eq{ \rho(\tau) \approx \rhocmbosc + \rhocmbnoise + \rhonoisenoise }
and
\eq{ R(\tau) \approx \Rcmbcmb + \Rnoisenoise . }
Since $\Diosct = f(t) \ricmbt$, we expect $\Diosctau \approx \ftau \ricmbtau$ and, therefore, $\rhocmbosc \approx \ftau \Rcmbcmb$. Then our estimator returns
\eq{ \hat{f}(\tau) \approx \frac{\ftau \Rcmbcmb + \rhocmbnoise + \rhonoisenoise}{\Rcmbcmb + \Rnoisenoise} . \label{eq:festimatorApproxExpression} }
The main non-idealities are due to noise. In the numerator, the high-variance but mean-zero correlations~$\rhocmbnoise$ and~$\rhonoisenoise$ cause a large scatter in measurements of~$\hat{f}(\tau)$. In the denominator, the noise autocorrelation~$\Rnoisenoise$, which is due to residuals in the coadded maps, causes an overall suppression. In the limit of negligible noise, we find $\hat{f}(\tau) \approx \ftau$.

%%%
\subsection{Signal transfer function \label{sec:sigTF}}

For a single season of data from the \emph{Keck Array}, the most significant bias in~$\hat{f}(\tau)$ comes from residual noise in the coadded maps. As shown in \eqref{eq:festimatorApproxExpression}, the bias tends to suppress the signal strength. For one season, the suppression is at the level of~$\sim 30\%$ with percent-level variation across observations. By constructing coadded maps from multiple seasons of \emph{Keck Array} observations, this effect can be reduced substantially in future analyses. 

For oscillation periods comparable to the length of a scanset, we expect a signal suppression similar to an averaging of the oscillation as in \eqref{eq:avgfsinc}. Instantaneously, we have $\Diosct = f(t) \Dircmbt$. If $f(t)$~is changing over the course of the scanset, then the binning, weighting and filtering required to produce~$\Diosctau$ will act on both~$f(t)$ and~$\Dircmbt$, i.e., we cannot treat~$f(t)$ merely as an overall constant scaling. The analysis, however, depends only on the correlation~$\rhosrosc$, which involves averaging over thousands of detectors and map pixels. When the oscillation period is much longer than the scanset, we can treat~$f(t)$ as approximately constant and write 
\eq{ \rhosrosc \approx f(\tau) \rhosrrcmb , }
where $\rhosrrcmb$, as defined in \eqref{eq:rhosrsD}, is the correlation between~$\risrtau$ and the unphysical but well-defined~$\Dircmbtau$. For shorter oscillation periods, we can subdivide the scanset into shorter time bins until the approximation is valid. Then the scanset-level correlation~$\rhosrosc$ is a weighted average of the correlations from the subdivisions. As the time-weighting is approximately uniform in BICEP observations that pass standard selection criteria, we approximate the weighted average with the uniform-weight average~$\ftau$ from \eqref{eq:avgfsinc} and write
\eq{ \rhosrosc \approx \ftau \rhosrrcmb . \label{eq:rhoRelatedToAvgfrhorCMB} }
This averaging suppresses the signal for oscillation periods on the order of a scanset. The suppression could be reduced by binning more finely in time. We argued in \secref{sec:timeBinning} that finer time binning is unmotivated given the constraints set by SN1987A.

%%%
\subsection{Simulations \label{sec:sims}}

The analysis has been cast in terms of the correlation quantities~$\rho(\tau)$ and~$R(\tau)$. We showed in \secref{sec:corrMatrices} how each correlation can be decomposed as a matrix, each element of which gives the correlation between two contributing components of the polarization field. The components we consider are static CMB, static foregrounds, noise and polarization oscillations. The first three are standard simulation products and are independent. The oscillations are derived from the static CMB according to \eqref{eq:QUmixingmatrix} and depend on three parameters: amplitude~$A$, phase~$\alpha$ and angular frequency~$m$, i.e., the input oscillation is parameterized as
\eq{ f(t) = A \cos(mt + \alpha) , \label{eq:f(t)parameterization} }
where, by assumption, $A \ll 1$.

Simulations are constructed from \emph{reobservations} of input maps using the real detector pointings, data cuts and weighting. For the static CMB maps, we use simulations of lensed $\Lambda\mathrm{CDM}$ cosmology, since that is the closest representation to the foreground-cleaned sky signal. Consistent with the standard BICEP simulation pipeline, we use Gaussian dust to simulate Galactic foregrounds. 

For noise, we use the real data with random sign flips assigned at each scanset and with the average value subtracted out. This ensures that the coadded noise-simulation map has the same noise properties as the real data, including detector covariances, with negligible residuals from CMB and foregrounds. By flipping signs randomly, any true oscillation signal is scrambled. Furthermore, the template in simulation is constructed from a different CMB realization than the real data, so the correlation method would not pick up an oscillation in the sign-flip noise, even if one were present. A sign-flip noise scheme is computationally efficient, since the noise realizations are simply drawn from the real data rather than from an additional set of reobservations.

For computational efficiency, a major goal of the analysis is to minimize the number of independent reobservations. In the \emph{Keck Array}, each scanset corresponds to roughly 50~minutes of observation time. In a full year of data, there are $\sim 4 \times 10^3$~scansets. Many of the scansets are approximately identical in terms of detector pointings, so signal-only simulations need only be run on a minimal set of independent observations, which typically consists of only $\sim 200$~scansets.

Sampling the oscillation parameters~$A$, $\alpha$ and~$m$ would be computationally infeasible through reobservations. Instead, we note from \eqref{eq:rhoRelatedToAvgfrhorCMB} that $\rhosrosc$~is always proportional to~$\rhosrrcmb$, which depends only on the input static CMB field and the detector pointing. We can then reobserve the unphysical rotated CMB map to create~$\Dircmbtau$ for each scanset and then correlate with~$\risrtau$ to construct~$\rhosrrcmb$. This need only be done once per realization. For the physical oscillation signal, we need only scale the result by~$\ftau$, which is related to the oscillation parameters through Eqs.~\ref{eq:avgfsinc} and~\ref{eq:f(t)parameterization}.

For the results presented below, we reobserved 110~realizations of lensed $\Lambda\mathrm{CDM}$ CMB, Gaussian dust and the rotated version of lensed $\Lambda\mathrm{CDM}$ CMB. For the noise simulations, we use 110~independent sign-flip sequences on the real data. For each realization, we compute the elements of the correlation matrices~$\rhosrsD$ and~$\Rsrsr$, where $s_r \in \{\mathrm{CMB},\mathrm{fg},\mathrm{N}\}$ and $s_D \in \{\mathrm{CMB},\mathrm{fg},\mathrm{N},\mathrm{rCMB}\}$. The efficiency is increased further by noting that $\Rsrsr$~is symmetric.

The correlations are saved to disk for each detector pair independently, so we can apply detector selections without redundant computation. We define the per-detector correlation matrix elements
\eq{ \rhoisrsD \equiv \frac{\sum\limits_{\unitVec{n}} \risrtau \DisDtau \witau{} \vitau{} }{W_i(\tau)} \label{eq:rhoisrsD def} }
and
\eq{ \Risrsr \equiv \frac{ \sum\limits_{\unitVec{n}} \risronetau \risrtwotau \witau{} \vitau{} }{W_i(\tau)} , \label{eq:Risrsr def} }
where
\eq{ W_i(\tau) \equiv \sum_{\unitVec{n}} \witau{} \vitau{} . \label{eq:WiDef} }
With these definitions, we can construct the all-detector matrix elements with
\eq{ \rhosrsD = \frac{1}{W(\tau)} \sum_i \rhoisrsD W_i(\tau) , \label{eq:rhosrsDrelatedtorhoisrsD} }
\eq{ \Rsrsr = \frac{1}{W(\tau)} \sum_i \Risrsr W_i(\tau) \label{eq:RsrsrrelatedtoRisrsr} }
and
\eq{ W(\tau) = \sum_i W_i(\tau) . }
To accommodate the scan-direction jackknife test described below in \secref{sec:systematics}, we separate the results for left- and right-going scans. We save the per-detector correlation to disk in this separated form and combine scan directions in a weighted average only once a jackknife test has been chosen that does not depend on scan direction.

\begin{figure*}
	\includegraphics[width = \textwidth]{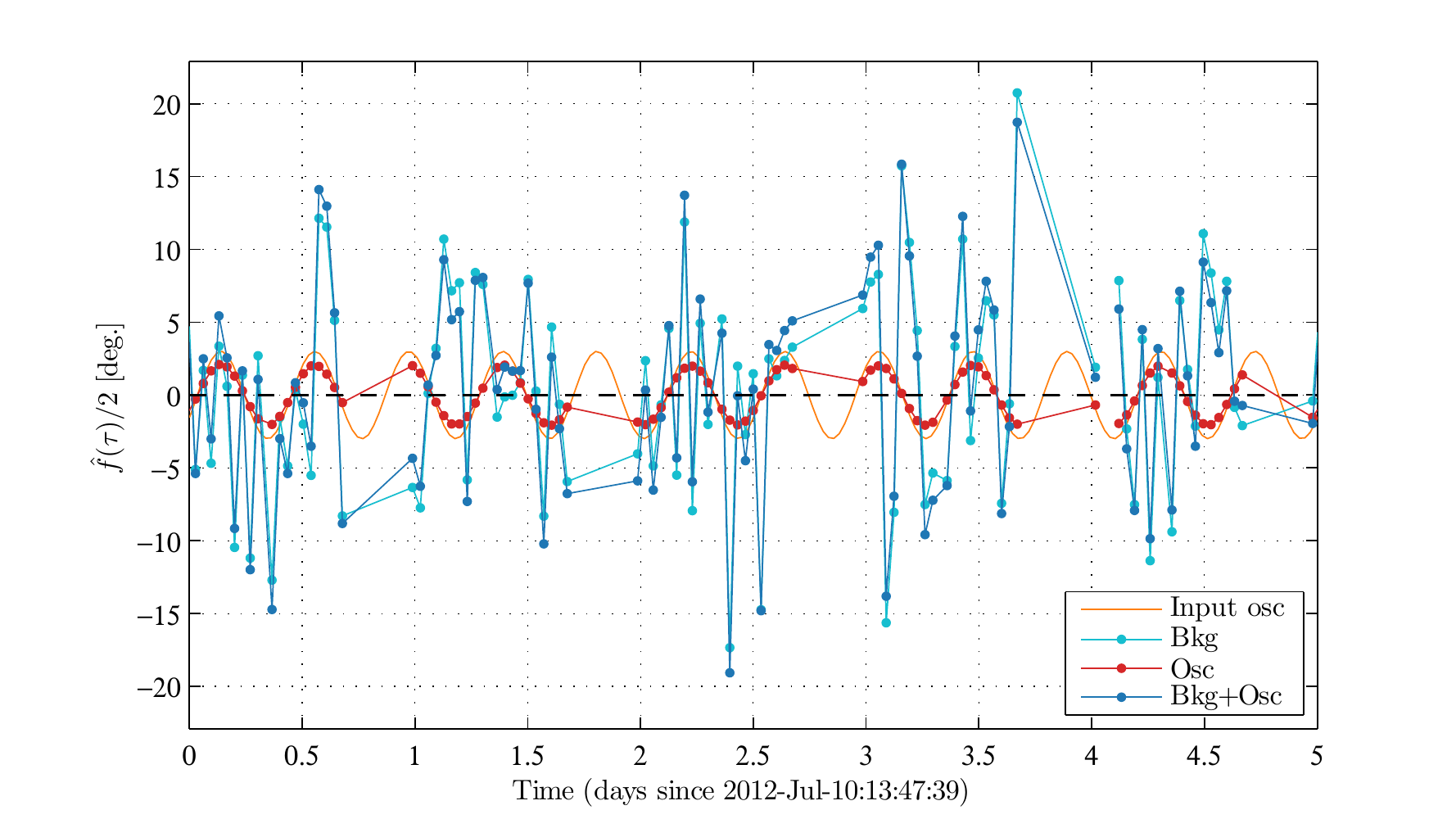}
	\caption{A simulated time series~$\hat{f}(\tau)$ (blue) with an input rotation amplitude $A/2 = 3^\circ$, chosen to be relatively large in order to illustrate the effect more clearly. The background~$\fhatbkg$ (cyan, defined explicitly in \eqref{eq:fhatbkgDef}) dominates over the oscillating component~$\fhatosc$ (red, \eqref{eq:fhatoscDef}). The template maps~$\Qcoadd{}$ and~$\Ucoadd{}$ are constructed from standard BICEP simulations of only the 2012 observing season of the \emph{Keck Array}, so the underlying true oscillation~$f(t)$ (orange) is not recovered at full strength but is instead suppressed by~$\sim 30\%$ as described in \secref{sec:sigTF}. \label{fig:f_example} }
\end{figure*}

An example of a simulated time series is shown in \figref{fig:f_example}, where we also isolate the contributions from background and from axion-like polarization oscillations.
The data are dominated by background fluctuations, and the oscillation is a small perturbation. As discussed in \secref{sec:sigTF}, the estimator~$\hat{f}(\tau)$ returns a slightly suppressed version of the true signal~$f(t)$.

The template maps~$\Qcoadd{}$ and~$\Ucoadd{}$ are constructed from all detectors observing at the same photon frequency. For this reason, the detector-related systematics tests described in \secref{sec:systematics} are \emph{partial} jackknives, since the template is constructed from \emph{all} detectors but the correlation sums in Eqs.~\ref{eq:rhoDef} and~$\ref{eq:Rdef}$ cover only \emph{half} of the detectors.

%%%
\section{Likelihood and Bayesian upper limits \label{sec:likelihood} }

We compare the data~$\hat{f}(\tau)$ to a model consisting of static CMB, static foregrounds, noise and a single oscillating component, i.e., we assume there is only one axion mass~$m$. From simulation, we construct the model distribution and estimate a likelihood for each candidate value of~$m$ independently. By imposing prior distributions on the amplitude~$A$ and phase~$\alpha$, we can set Bayesian upper limits on the axion-photon coupling constant, which is directly related to the oscillation amplitude~$A$. At the same time, we fit for amplitude and phase and form a test statistic to check for consistency with the background-only model, where we take the background to consist of static CMB, static foregrounds and noise.

\subsection{Model distribution from simulation \label{sec:modelDist}}

In simulation, we can decompose the estimator~$\hat{f}(\tau)$ (\eqref{eq:fhatDef}) as a matrix by
\eq{ \fhatsrsD \equiv \frac{\rhosrsD}{R(\tau)} , }
where $\rhosrsD$~is defined in \eqref{eq:rhosrsD} and $R(\tau)$~in \eqref{eq:Rdef}. The denominator, then, contains contributions from all of the non-oscillating map components. From \eqref{eq:rhoRelatedToAvgfrhorCMB}, we can pull out the dependence on~$\ftau$ and rewrite the oscillating elements in terms of the rotated map, i.e., 
\eq{ \fhatsrosc = \ftau \fhatsrrcmb . }
This is a convenient factorization, since we save~$\rhosrrcmb$ to disk but not~$\rhosrosc$ as described in \secref{sec:sims}. We define the \emph{dynamic} mixing angle
\eq{ \fhatdyn \equiv \sum_{s_r} \fhatsrrcmb , }
where $s_r \in \{\mathrm{CMB},\mathrm{fg},\mathrm{N}\}$. 
The oscillating component of the mixing angle is, then, 
\eq{ \fhatosc \equiv \ftau \fhatdyn . \label{eq:fhatoscDef} }
Independent of any particular choice for oscillation parameters~$A$, $\alpha$ and~$m$, we can use the dynamic mixing angle~$\fhatdyn$ to investigate and precompute statistics of the signal transfer function.
We define the \emph{background} mixing angle by
\eq{ \fhatbkg \equiv \sum_{s_r,s_D} \fhatsrsD , \label{eq:fhatbkgDef} }
where $s_r,s_D \in \{\mathrm{CMB},\mathrm{fg},\mathrm{N}\}$. Then the full mixing angle is
\spliteq{ \hat{f}(\tau) & = \fhatbkg + \fhatosc \\
	& = \fhatbkg + \ftau \fhatdyn . }

We simulate many realizations of~$\hat{f}(\tau)$. The mean over realizations is
\eq{ \fhatavg = \avg{ \fhatbkg } + \ftau \fhatdynavg , }
so we need save to disk only~$\fhatbkgavg$ and~$\fhatdynavg$, neither of which depends on oscillation parameters. To explore the oscillation parameter space, we scale the dynamic component by~$\ftau$ and take the linear combination with the background component. In the limit of noiseless coadded maps, the dynamic component is close to unity. From \eqref{eq:festimatorApproxExpression}, however, we see that residual noise in the coadded maps will suppress the signal. The quantity~$\fhatdynavg$ is a measure of this suppression. We find $\fhatdynavg \approx 70\%$ with one season of $150$-$\mathrm{GHz}$ data from the \emph{Keck Array}, and the variance over~$\tau$ is at the percent level. The variance in~$\fhatdyn$ over realizations for a given scanset~$\tau$ is also at the percent level. In general, the consequence of residual noise is a consistent suppression of the signal by roughly~$30\%$. The suppression can be lessened by using more seasons of data to form the coadded maps.

By assumption, the true background mean vanishes. Even though the sample mean over realizations may be non-zero, we set the background model mean to zero. Then the full model mean, i.e., including oscillations, is
\eq{ \mu(\tau) \equiv \ftau \fhatdynavg . \label{eq:modelMean} }
The variance over realizations is dominated by the background, so we take the model variance to be
\eq{ \sigma^2(\tau) \equiv \avg{ \parens{ \fhatbkg }^2 } . \label{eq:modelStd} }
With the ensemble of simulations, one can check that the standardized variable 
\eq{ s_\tau \equiv \frac{ \hat{f}(\tau) - \mu(\tau) }{\sigma(\tau)} }
is Gaussian distributed with mean zero and unit variance when $\hat{f}(\tau)$~is created with the same oscillation parameters that define~$\mu(\tau)$. Additionally, no evidence is found for covariances between scansets. We take it as a model assumption, then, that $s_\tau$~is drawn independently for each scanset time~$\tau$ from a standard Gaussian distribution.

%%%
\subsubsection{Pseudo-simulations \label{sec:pseudosims}}

We can quickly form pseudo-simulations of the time series~$\hat{f}(\tau)$ by resampling from a Gaussian distribution with mean~$\mu(\tau)$ (\eqref{eq:modelMean}) and standard deviation~$\sigma(\tau)$ (\eqref{eq:modelStd}). In this way, we can avoid the computational expense of a large number of reobservations. In the results presented below, we use 110~reobservations to estimate~$\mu(\tau)$ and~$\sigma(\tau)$ for each scanset time~$\tau$. To estimate $p$-values, however, we fill out the distribution with a larger number of pseudo-simulations: $2 \times 10^4$~realizations to test for consistency with the background model and $5 \times 10^3$~realizations to test for spurious systematic excesses.

%%%
\subsection{Likelihood function \label{sec:likelihoodFunction}}

We form a Gaussian likelihood for a three-parameter oscillation model. Let $\mu(\tau;m,A,\alpha)$ be the model mean (\eqref{eq:modelMean}) formed when $f(t) = A \cos(mt + \alpha)$. Then we form the test statistic
\eq{ q_m(A,\alpha) \equiv \sum_\tau \brackets{ \frac{ \hat{f}(\tau) - \mu(\tau;m,A,\alpha) }{\sigma(\tau)} }^2 , \label{eq:qmDef} }
where the estimator~$\hat{f}(\tau)$ is formed from the input data.
We remove from the analysis any scansets whose model variance~$\sigma^2(\tau)$ is more than two standard deviations from the mean model variance. The associated observation times~$\tau$ simply do not contribute to the sum in \eqref{eq:qmDef}. As the model is assumed to be Gaussian, the quantity~$q_m(A,\alpha)$ is a $\chi^2$~test statistic with $n$ degrees of freedom, where $n$~is the number of scansets contributing to the sum.

In general, we will consider each value of~$m$ separately. For each~$m$, we will allow amplitude~$A$ and phase~$\alpha$ to vary, so we form an ensemble of likelihoods, one for each value of~$m$:
\eq{ L_m(A,\alpha) = N \exp \brackets{ - \frac{q_m(A,\alpha)}{2} } , }
where
\eq{ N \equiv \frac{1}{ \sqrt{ (2 \pi)^n \prod_\tau \sigma^2(\tau) } } }
is a normalization coefficient. Crucially, there is no dependence on oscillation parameters in~$N$, so it will be convenient to consider likelihood \emph{ratios}, for which the $N$-dependence drops out.

For computational efficiency, it is convenient to expand the test statistic~$q_m(A,\alpha)$ from \eqref{eq:qmDef} as a linear combination of $\tau$-sums, each of which depends only on~$m$ and \emph{not} on~$A$ nor~$\alpha$. These terms can be evaluated for a chosen set of $m$-values and saved to disk. The two-dimensional parameter space of amplitude~$A$ and phase~$\alpha$ can then be explored quickly by forming linear combinations of the $m$-dependent terms.

%%%
\subsection{Bayesian upper limits \label{sec:BayesianUpperLimits}}

Having computed the likelihood~$L_m(A,\alpha)$, we marginalize over the phase~$\alpha$, which carries no information about axion properties and is expected to be random. We set a uniform prior on the phase and define the marginal likelihood
\eq{ L_m(A) \equiv \frac{1}{2 \pi} \int_0^{2\pi} d\alpha~L_m(A,\alpha) . }
For the amplitude~$A$, we impose a prior distribution~$P(A)$. The prior could, in principle, vary with~$m$. For example, the prior could incorporate axion constraints from other datasets. For this analysis, however, we set a uniform, $m$-independent prior.

The posterior distribution for each $m$-value is
\eq{ P_m\parens{A \middle | \squiggles{\hat{f}(\tau)} } = \frac{ P(A) L_m(A) }{\int dA~P(A) L_m(A)} , }
i.e., the probability density for amplitude~$A$ given the angular frequency~$m$ and the data~$\squiggles{\hat{f}(\tau)}$. We set a linearly uniform prior
\eq{ P(A) = \frac{1}{A_\mathrm{max}} \brackets{0 \leq A \leq A_\mathrm{max}}, }
where $A_\mathrm{max}$ is set conservatively above the current constraints in the axion mass range under consideration. In the results below, we have used $A_\mathrm{max}/2 = 4^\circ$.\footnote{We will often be interested in the quantity~$A/2$, since it is the amplitude of the on-sky oscillation of polarization angles. The quantity~$A$ is the amplitude of the mixing of Stokes~$Q$ and~$U$.} We integrate the posterior to estimate a cumulative distribution function (CDF), which we again compute for each $m$-value independently:
\eq{ F_m(A) = \int_0^A dA'~P_m\parens{A' \middle | \squiggles{\hat{f}(\tau)}} . }

We set a Bayesian 95\% credible interval by finding the amplitude~$A$ that satisfies the condition $F_m(A) = 95\%$. This is our upper limit on the oscillation amplitude at each $m$-value.

%%%
\subsection{Background consistency \label{sec:bkgConsistencyTest}}

To check for consistency with the background, we form a $\Delta\chi^2$~test statistic. First we evaluate the test statistic (\eqref{eq:qmDef}) for the background model, i.e., with $A = 0$:
\eq{ q_0 \equiv q_m(0,\alpha) . \label{eq:q0Def} }
Then we find the amplitude~$\hat{A}_m$ and phase~$\hat{\alpha}_m$ that minimize~$q_m(A,\alpha)$. The test statistic for background consistency is
\eq{ \Delta q_m \equiv q_0 - q_m\parens{\hat{A}_m,\hat{\alpha}_m} , \label{eq:deltaqmDef} }
which is expected to be $\chi^2$-distributed with 2~degrees of freedom. This expectation is confirmed in simulations, but we do not rely on it in the results presented below. Instead, all $p$-values are calibrated with an ensemble of simulations.

We evaluate the test statistic~$\Delta q_m$ from \eqref{eq:deltaqmDef} for each $m$-value. In the results presented below, we consider $\sim 10^4$~values, so it is necessary to account for a trials factor. We use 
\eq{ \Delta\hat{q} \equiv \max_{m>0} \parens{ \Delta q_m } \label{eq:deltaqhatDef} }
as a global test statistic. We exclude $m = 0$ from the maximization, since this term represents a constant offset and not an oscillation. Let~$\hat{p}$ be the associated probability-to-exceed (PTE) or $p$-value for~$\Delta \hat{q}$. In the results presented below, we have calibrated~$\hat{p}$ with $2 \times 10^4$~realizations and can, therefore, estimate statistical tension up to the level of~$\sim 3\sigma$.

%%%
\section{Systematics \label{sec:systematics}}

We perform a set of data splits to test for systematic effects that could create spurious oscillation signals. The set of splits is identical to those performed in previous BICEP/\emph{Keck Array} analyses (cf.~\cite{BK-III}), but the implementation and interpretation are different.

The template maps~$\Qcoadd{}$ and~$\Ucoadd{}$ are constructed from the full dataset regardless of which jackknife test is under consideration. The rotated map~$\ritau{}$ (defined in \eqref{eq:ritauDef}) depends on the orientation of detector~$i$ at scanset time~$\tau$ but is otherwise simply a linear combination of the template maps. For each scanset~$\tau$, we form the per-detector correlation matrix elements~$\rhoisrsD$ and~$\Risrsr$ defined in Eqs.~\ref{eq:rhoisrsD def} and~\ref{eq:Risrsr def}, respectively. The correlations are constructed separately for left- and right-going scans, of which there are approximately $50$~each per scanset. Having saved to disk the per-detector correlations split by scan direction for each scanset, we can perform jackknife tests by selecting subsets of these quantities to form a time series~$\hat{f}(\tau)$ and a likelihood~$L_m(A)$. For the nontemporal jackknife tests, we form the mixing-angle estimator~$\hat{f}(\tau)$ for each scanset~$\tau$ from half of the data, either from only one of the scan directions or from only half of the detectors. For the temporal jackknife tests, we form the mixing-angle estimator~$\hat{f}(\tau)$ from all of the data available at each scanset, but we form the likelihood~$L_m(A)$ from only half of the scansets.

%%%
\subsection{Nontemporal jackknife tests \label{sec:nontemporalJackknives}}

The nontemporal jackknives test whether the estimator~$\hat{f}(\tau)$ is a good statistical representation of the data collected during scanset~$\tau$. We split the data either by the direction the telescope is slewing or by the contributing detectors, and we search for a systematic difference in the results. 
We perform the following nine nontemporal data splits defined in Sec.~8 of~\cite{BK-III}: \emph{Scan direction}, \emph{Tile}, \emph{Tile/deck}, \emph{Focal plane inner/outer}, \emph{Tile top/bottom}, \emph{Tile inner/outer}, \emph{Mux column}, \emph{Mux row} and \emph{Differential pointing best/worst}. The scan-direction jackknife is considered nontemporal for the axion-oscillation analysis, since the left- and right-going scans are interleaved on timescales much smaller than the oscillation periods of interest. The other jackknives split the data to expose potential non-idealities in optical, detector or readout properties.

For these jackknife tests, it is possible to cancel the time-domain signal at the scanset level. We form a time series for the jackknife difference
\eq{ \fjk \equiv \frac{\hat{f}^{(1)}(\tau) - \hat{f}^{(2)}(\tau)}{2} , }
where $\hat{f}^{(i)}(\tau)$~is the mixing-angle estimator formed from the $i$th~half of the data split.
We can treat~$\fjk$ as an ordinary time series.
In simulation, one can check the efficacy of the signal cancellation by looking at~$\fhatdynavg$ (introduced in \secref{sec:modelDist}) constructed from many realizations of~$\fjk$. Recall that $\fhatdynavg \sim 70\%$ for undifferenced data. With differencing, there is a substantial reduction in~$\fhatdynavg$, though there is variation among the jackknife tests. As the noise level is expected to be comparable to current limits on the axion-photon coupling constant, we need only require a relatively modest signal cancellation. All of the tests reduce~$\fhatdynavg$ by more than a factor of~$20$ with percent-level scatter over~$\tau$. The scan-direction jackknife cancels the signal significantly better than all other tests, while the \emph{Focal plane inner/outer} test cancels worst with a few-percent bias away from zero.

Treating~$\fjk$ as a measured rotation-angle time series, we can evaluate the test statistic (\eqref{eq:deltaqmDef}) for background consistency:
\eq{ \Delta q_m^{(\mathrm{jk})} \equiv \Delta q_m \parens{ \squiggles{ \hat{f}^{(\mathrm{jk})}(\tau) } } . }
Even though we expect any time-variable signal to be cancelled, we must perform this consistency test as a function of~$m$. If we only used, e.g., the test statistic~$q_0$ (\eqref{eq:q0Def}) to check for background consistency, we would not pick up small residual oscillations. Sinusoidal fitting has much greater sensitivity to oscillatory signals, so we use the test statistic~$\Delta q_m$ (\eqref{eq:deltaqmDef}) for these jackknife tests as well. The $\Delta q_m$ test statistic is formed by comparing to the \emph{undifferenced} model distribution. We do not compare to a model distribution based on the jackknife difference, because the signal transfer function for oscillations is close to zero.
If~$\hat{f}^{(1)}(\tau)$ and $\hat{f}^{(2)}(\tau)$~have equal variances, then $\fjk$~has the same variance as the background. In that case, $\Delta q_m^{(\mathrm{jk})}$~follows a $\chi^2$-distribution with 2~degrees of freedom just like~$\Delta q_m$ from \eqref{eq:deltaqmDef}. In general, the variances are \emph{not} equal, but the test statistic~$\Delta q_m^{(\mathrm{jk})}$ can be scaled by an $\mathcal{O}(1)$~factor, which can be fit for, to map it onto a $\chi^2$-distribution with 2~degrees of freedom. We do not, however, rely on the $\chi^2$~expectation for any results and instead calibrate $p$-values for~$\deltaqmjk$ through simulation.

Just as discussed for the background-consistency test in \secref{sec:bkgConsistencyTest}, we must account for the large number of $m$-values being tested. We define
\eq{ \Delta\hat{q}^{(\mathrm{jk})} \equiv \max_{m > 0} \parens{ \Delta q_m^{(\mathrm{jk})} } , \label{eq:deltaqjkDef} }
the most extreme signal-like excess, as a global test statistic for consistency with simulations. We exclude $m = 0$ from the maximization, since this value represents a constant rotational offset, which may indicate an inefficiency in the experiment but does not produce spurious signals. We estimate a $p$-value for this test statistic by comparing to a distribution of background-only pseudo-simulations differenced in the same manner. Since the signal mostly cancels, the background-only simulations give approximately the same results as simulations with signal included. 

%%%
\subsection{Temporal jackknife tests \label{sec:temporalJackknives}}

Whereas the nontemporal jackknives test for statistical consistency \emph{within} a scanset, the temporal jackknives test for consistency \emph{among} scansets. We split the scansets into two groups and search for a systematic difference in oscillation signals. 
We perform the following five temporal data splits defined in Sec.~8 of~\cite{BK-III}: \emph{Deck angle}, \emph{Alternative deck}, \emph{Temporal split}, \emph{Azimuth} and \emph{Moon up/down}. These tests are designed to expose pick-up from far sidelobes and non-idealities of the optical performance. Because we are searching for a time-varying signal, it is not possible for temporal jackknives to cancel the signal in the time domain. Instead, we cancel the signal in the frequency domain by considering the best-fit oscillation amplitude constructed from each half of the temporal split. We form the test statistic
\eq{ \hat{A}_m^{(\mathrm{jk})} \equiv \left | \frac{ \hat{A}_m^{(1)} - \hat{A}_m^{(2)}}{2} \right | , \label{eq:AjkDef} }
where $\hat{A}_m^{(i)}$~is the best-fit amplitude formed from the $i$th~half of the data split. We take the absolute value of the difference, so signal-like systematics appear on only one side of the test-statistic distribution. Since the temporal data splits impose different window functions on the time series~$\hat{f}(\tau)$, the best-fit amplitudes may vary even when the true frequency content is the same. This is similar to apodization effects in Fourier transforms, though we are not computing a Fourier transform here. This test statistic cancels the signal by more than an order of magnitude, but the residual has a larger variance than the background. We accept this increased variance from potential signals but still require the real jackknife results to match a background-only model. The test statistic~$\hat{A}_m^{(\mathrm{jk})}$ is distributed approximately as a one-sided Gaussian, but we estimate all $p$-values by comparing with simulations. 

Due to the signal transfer function from time binning, the variance in~$\Ajk$ increases with~$m$. To keep the same normalization across the entire $m$-range, we divide by the standard deviation as measured from background-only simulations. The test statistic we use for estimating statistical deviations from the model distribution is, then,
\eq{ a^{(\mathrm{jk})}_m \equiv \frac{\Ajk}{\operatorname{std} \brackets{\Ajk}} . }

As in Secs.~\ref{sec:bkgConsistencyTest} and~\ref{sec:nontemporalJackknives}, we must account for the large number of $m$-values under consideration. We take the largest value over the $m$-range, which is the most extreme signal-like excess:
\eq{ \hat{a}^{(\mathrm{jk})} \equiv \max_{m > 0} \parens{ a_m^{(\mathrm{jk})} } . \label{eq:ajkDef} }
We exclude $m = 0$ from the maximization, since this term indicates a constant rotational offset, which produces spurious signals in a way that is dealt with in \secref{sec:tempJKconstOffset}.
We estimate a $p$-value for this test statistic by comparing to a distribution of background-only pseudo-simulations that have been subjected to the same temporal jackknife. Although more signal can leak through the temporal jackknives than the nontemporal jackknives, we require the real data to be statistically consistent with the background-only simulations. This is a stricter requirement than is necessary to test for spurious systematic signals.

\subsubsection{Constant offset \label{sec:tempJKconstOffset}}

The temporal jackknives test for two different types of systematics. First, we want to check that any oscillations are appearing at the same level in both halves of the temporal split. Second, we want to know if there is a systematic bias in the rotation angle that depends on a time-variable scan parameter and can, therefore, produce spurious oscillation signals at frequencies related to the observing schedule. The test statistic~$\hat{a}^{(\mathrm{jk})}$ addresses the first type of concern. The second concern is addressed by privileging the $m = 0$ jackknife difference, i.e., the difference in constant offset between the two halves of the data split. If there is a systematic bias that depends on, e.g., deck angle, then a naive analysis may detect an oscillation signal at frequencies related to the deck-rotation schedule. A jackknife split,  however, will discover the bias as a statistically anomalous value for~$\hat{A}^{(\mathrm{jk})}_0$. We, therefore, include~$\Adc$ as a test statistic for the \emph{Deck angle}, \emph{Alternative deck}, \emph{Azimuth} and \emph{Moon up/down} jackknife tests. The reason we omit the \emph{Temporal split} is that a non-zero value for~$\Adc$ could also be produced by a long-period oscillation. In fact, this jackknife test represents a minor unblinding, since a true oscillation, if it happens to be roughly synchronized with one of the temporal-jackknife timescales, could appear in~$\Adc$. Since this is unlikely, we proceed with the jackknife test and treat any deviation from the background-only simulations as evidence for a systematic bias.

%%%
\subsection{Global systematics assessment \label{sec:globalSysAssessment}}

For the nine nontemporal jackknives listed in \secref{sec:nontemporalJackknives}, we calculate the test statistic~$\deltaqjk$ (\eqref{eq:deltaqjkDef}). For the five temporal jackknives listed in \secref{sec:temporalJackknives}, we calculate the test statistic~$\ajk$ (\eqref{eq:ajkDef}). For the four temporal jackknives selected in \secref{sec:tempJKconstOffset}, we calculate the test statistic~$\Adc$ (\eqref{eq:AjkDef}). In total, then, we are performing 18~tests for consistency with simulations. For each test, we form a $p$-value, which we will denote~$p_i$, where $i$~is an index labeling each of the 18~tests. All of these $p$-values are calibrated by comparing with an ensemble of $5 \times 10^3$ background-only pseudo-simulations (\secref{sec:pseudosims}). A simulation ensemble of this size allows us to estimate $p$-values down to the level of~$\sim 10^{-3}$. It is not important to precisely estimate smaller values, since we consider values below this level to indicate unacceptable inconsistency with the model. If we obtain such extreme values, we would consider our measurements to be systematically biased and would investigate the source before unblinding the undifferenced data.

Because we perform 18~systematics tests, we must account for a trials factor in determining the statistical significance of the most extreme result. We take the minimum jackknife $p$-value
\eq{ \hat{p}^{(\mathrm{sys})} \equiv \min_i p_i \label{eq:hatpsys} }
as a global test statistic for consistency with the simulation ensemble. While $\hat{p}^{(\mathrm{sys})}$~tests for signal-like systematics, we also check the sensitivity of the jackknife tests with the test statistic
\eq{ \hat{c}^{(\mathrm{sys})} \equiv 1 - \max_i p_i , \label{eq:hatcsys} }
which is a measure of how well the model variances have been estimated. A small value for~$\hat{c}^{(\mathrm{sys})}$ indicates that the variances have been overestimated, which would degrade the sensitivity of the jackknife tests.

The quantities~$\hat{p}^{(\mathrm{sys})}$ and~$\hat{c}^{(\mathrm{sys})}$ are drawn from an ensemble of $p$-values but are to be regarded as test statistics. We use the ensemble of 110~reobservations to estimate $p$-values for these test statistics. The reason we do not use the ensemble of $5 \times 10^3$~pseudo-simulations is that these do not include covariances between jackknife tests. The reobservations show negligible covariances among the 18~test statistics, but we do not wish to depend on that statistical independence for any of our results. Let $p_p$~be the $p$-value of the test statistic~$\hat{p}^{(\mathrm{sys})}$ and $p_c$~the $p$-value for~$\hat{c}^{(\mathrm{sys})}$. Because these $p$-values are estimated from an ensemble of only 110~realizations, we provide only two significant figures instead of the three significant figures provided for the individual jackknife tests. 

We require both~$\hat{p}^{(\mathrm{sys})}$ and~$\hat{c}^{(\mathrm{sys})}$ to lie within the central $2\sigma$~region of the background distribution. Although there are two test statistics, we do not account for a trials factor. Our requirement, therefore, is more stringent than simply requiring overall $2\sigma$~consistency.

%%%
\section{Results \label{sec:results}}

For a first demonstration of the methods described in Secs.~\ref{sec:rotEst}, \ref{sec:likelihood} and~\ref{sec:systematics}, we selected the 2012 observing season of the \emph{Keck Array}. During this season, all five receivers observed at $150~\mathrm{GHz}$, and the dataset has been thoroughly vetted by the standard BICEP CMB analyses~\cite{BK-X}. The data volume is small enough for relatively quick iteration but large enough to understand computational scalings. The 2012 season represents only a small fraction of the total BICEP dataset, and we intend to extend this analysis to include more data in future publications.

An important element in the analysis is the rotated map~$\ritau{}$ (Eqs.~\ref{eq:ritDef} and~\ref{eq:ritauDef}), which is constructed from the coadded maps~$\Qcoadd{}$ and~$\Ucoadd{}$. In principle, these coadded maps could be constructed from the full BICEP dataset, while the time series~$\hat{f}(\tau)$ could be restricted to the 2012 season. For computational speed, however, we used only 2012 data in all components of the analysis, which produces a moderate but non-negligible signal suppression (\secref{sec:sigTF}).

%%%
\subsection{Mass coverage}

Our upper limits (\secref{sec:BayesianUpperLimits}) are estimated for each value of~$m$ independently. The set of $m$-values represents a discrete sampling in mass space rather than a binning. These $m$-values can, in principle, be chosen arbitrarily. We wish, however, to achieve approximately continuous coverage over as broad a mass range as possible. Unlike a discrete Fourier transform (DFT), we aim to have some redundancy between neighboring mass values in order to justify interpolation. The methods of Secs.~\ref{sec:likelihood} and~\ref{sec:systematics} do not require the results from different mass values to be independent.

Since we bin our results by $\sim45$-minute scansets (\secref{sec:timeBinning}), we take the minimum oscillation period considered in the analysis to be~$2~\mathrm{hr}$. As described in \secref{sec:timeBinning}, the constraints on~$g_{\phi\gamma}$ from SN1987A are sufficiently strong that there is little motivation to push to oscillation periods smaller than~$2~\mathrm{hr}$. This sets a maximum frequency for the analysis $\nu_\mathrm{max} = 0.5~\mathrm{hr}^{-1}$ and an associated maximum mass~$m_\mathrm{max} = 2 \pi \nu_\mathrm{max}$. Let~$T$ be the total time range covered by the time series~$\hat{f}(\tau)$. Each season, science observations for the \emph{Keck Array} typically lasted from early March until late October, so $T \approx 8~\mathrm{mo}$ for a single season. We set a frequency resolution $\Delta\nu \equiv 1/(\beta T)$, where $\beta$~is a factor that determines the amount of oversampling relative to a conventional DFT. In the results shown below, we use $\beta = 3$. With this frequency resolution, we consider the mass range $0 \leq m \leq  m_\mathrm{max}$. For the 2012 \emph{Keck Array} season, this amounts to 8638 $m$-values. The $m = 0$ results are used only for the temporal-jackknife test statistic~$\hat{A}_0^{(\mathrm{jk})}$ and are explicitly excluded from all other data products. We also ignore the results for oscillation periods longer than $30~\mathrm{d}$ in order to satisfy the approximation that the coadded maps contain only a negligible contribution from polarization oscillations (\eqref{eq:rOnly3components}). This last condition removes only 23 $m$-values, but it limits the low-frequency extent of our results. In a future iteration of the analysis, it may be computationally feasible to account for the oscillation residual in the coadded maps and set limits at arbitrarily low frequencies, though we expect degraded sensitivity when the oscillation period is on the order of or larger than the total observing time.

%%%
\subsection{Unblinding procedure \label{sec:unblinding}}

All real data products were kept blinded until the jackknife tests had been designed and shown in simulation to substantially suppress oscillation signals. From that point on, the results of real jackknife tests were unblinded. When it was concluded that there was no evidence for systematic effects in the jackknife tests, we agreed on a decision tree for unblinding the undifferenced data. Since the \emph{Keck Array} has collected data for eight seasons, the results from the 2012 season shown below represent a \emph{partial} unblinding of the full dataset. To prepare for the possibility of a signal-like excess, we decided \emph{before} unblinding that a measurement of $\hat{p} \leq 6.2 \times 10^{-3}$ (see \secref{sec:bkgConsistencyTest}), which would indicate tension with the background model in excess of~$2.5\sigma$, would trigger an analysis of an additional season of data and allow for unblinded investigation of systematic effects in 2012 data. If the excess persisted, it would trigger an analysis of all four of the seasons of \emph{Keck Array} observations contributing to~\cite{BK-X}. This strategy would allow us to distinguish between a real signal and a statistical fluctuation while also providing more opportunities and data to search for systematic effects. We measured $\hat{p} = 0.14$, which indicates $1.1\sigma$~signal-like tension with the background model. As this tension is significantly below the $2.5\sigma$~threshold, we present results from only the 2012 season below. The results from additional seasons are being processed, and we intend to present them in subsequent publications.

%%%
\subsection{Systematics}

The results of the 18~jackknife tests described in \secref{sec:systematics} are shown in Tabs.~\ref{tab:nontemporalJKtests} and~\ref{tab:temporalJKtests}.
\begin{table}
\begin{center}
\begin{ruledtabular}
\begin{tabular}{lc}
Jackknife test & $p\parens{\Delta \hat{q}^{(\mathrm{jk})}}$ \\
\hline
Scan direction &  $0.954$ \\
Tile & 0.709 \\
Tile/deck & 0.333 \\
Focal plane inner/outer & 0.434 \\
Tile top/bottom & 0.965 \\
Tile inner/outer & 0.534 \\
Mux column & 0.970 \\
Mux row & 0.994 \\
Differential pointing best/worst & 0.999
\end{tabular}
\end{ruledtabular}
\end{center}
\caption{Nontemporal-jackknife $p$-values for the test statistic~$\Delta\hat{q}^{(\mathrm{jk})}$ (\eqref{eq:deltaqjkDef}). Several $p$-values are close to~$1$, and this may be due to overestimates of the model variances, which would weaken but not invalidate the jackknife tests. We assess the statistical significance of the most extreme value with the test statistic~$\hat{c}^{(\mathrm{sys})}$ (\eqref{eq:hatcsys}), which gives a $p$-value $p_c = 0.044$ and is, therefore, in tension with the background model at the level of~$1.7\sigma$.  \label{tab:nontemporalJKtests}}
\end{table}
\begin{table}
\begin{center}
\begin{ruledtabular}
\begin{tabular}{lcc}
Jackknife test & $p\parens{\hat{a}^{(\mathrm{jk})}}$ & $p\parens{\hat{A}^{(\mathrm{jk})}_0}$ \\
\hline
Deck angle & 0.481 & 0.220 \\
Alternative deck & 0.330 & 0.621 \\
Temporal split & 0.127 & -- \\
Azimuth & 0.760 & 0.927 \\
Moon up/down & 0.621 & 0.191
\end{tabular}
\end{ruledtabular}
\end{center}
\caption{Temporal-jackknife $p$-values for the test statistics~$\hat{a}^{(\mathrm{jk})}$ (\eqref{eq:ajkDef}) and~$\hat{A}^{(\mathrm{jk})}_0$ (\eqref{eq:AjkDef} and \secref{sec:tempJKconstOffset}). As discussed in \secref{sec:tempJKconstOffset}, we do not consider~$\hat{A}^{(\mathrm{jk})}_0$ for the temporal split. \label{tab:temporalJKtests}}
\end{table}
We find $p_i \geq 12.7\%$ for all tests~$i$, and this indicates that no individual test has revealed a spurious signal. Two of the nontemporal jackknife tests (\tabref{tab:nontemporalJKtests}) show $p_i > 99\%$, which may be interpreted as statistically anomalous agreement with the background model. These large $p$-values suggest possible overestimates of the model variances, which lower the sensitivity of the jackknife tests to systematic effects. We provide a global assessment of the jackknife results by considering the test statistics~$\hat{p}^{(\mathrm{sys})}$ (\eqref{eq:hatpsys}) and~$\hat{c}^{(\mathrm{sys})}$ (\eqref{eq:hatcsys}). The associated $p$-values come to $p_p = 0.45$ and $p_c = 0.044$, respectively. The latter value is relatively low but lies within the central $2\sigma$~region ($0.0228 \leq p \leq 0.9772$), and we, therefore, conclude that there is no statistically significant tension with the background model.

A possible improvement for a future iteration of the systematics analysis is to consider, in addition to the most extreme $p$-values~$\hat{p}^{(\mathrm{sys})}$ and~$\hat{c}^{(\mathrm{sys})}$, the $p$-value \emph{distribution}, though it should be noted that the introduction of additional test statistics dilutes the sensitivity of each.

%%%
\subsection{Background consistency \label{sec:bkgConsistencyResults}}

\begin{figure}
	\includegraphics[width = 0.5\textwidth]{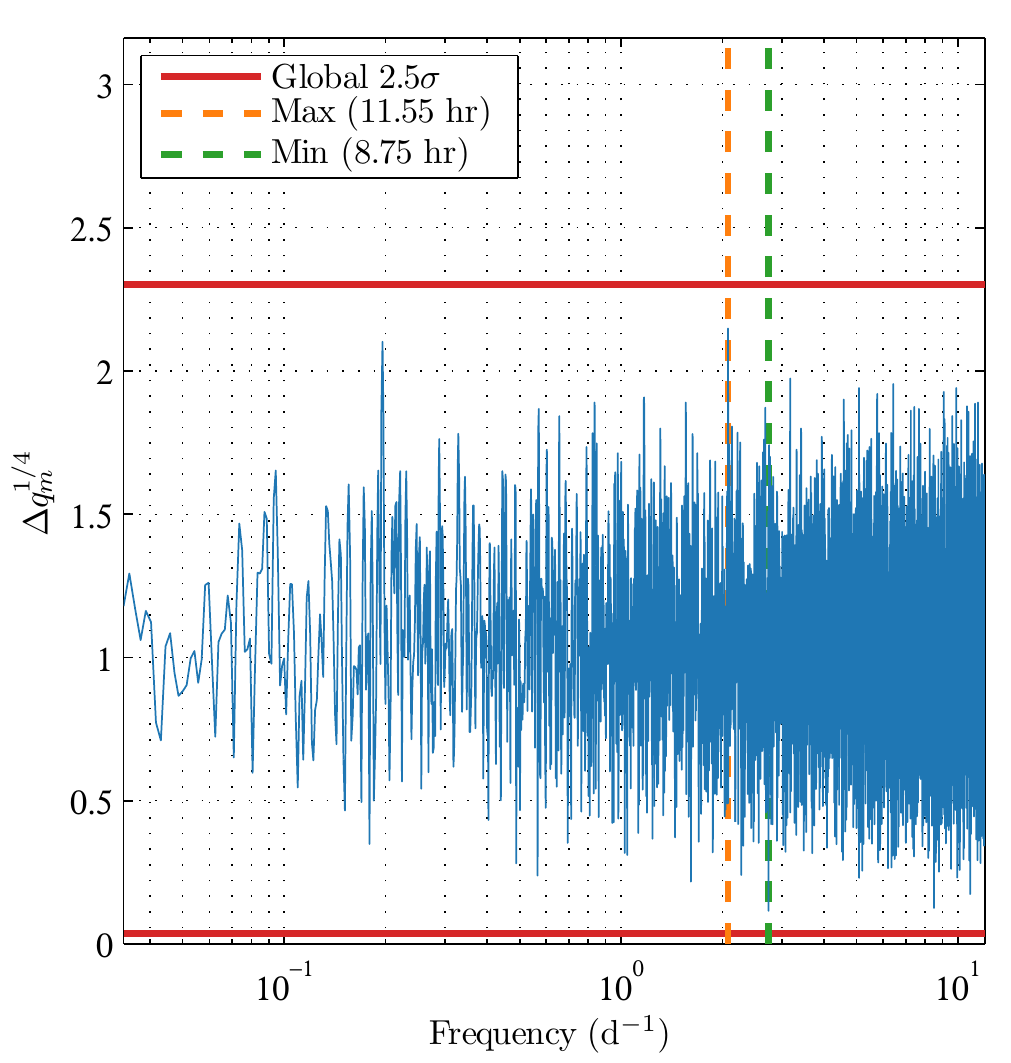}
	\caption{The test statistic~$\Delta q_m$ for consistency with the background model (\eqref{eq:deltaqmDef}) for real data from the 2012 observing season of the \emph{Keck Array}. We plot~$\Delta q_m^{1/4}$ on the vertical axis in order to compress the distribution for visual purposes, and we plot frequency~$m/(2\pi)$ in units of inverse days~($\mathrm{d}^{-1}$) on the horizontal axis. The maximum and minimum values are indicated in the legend with their corresponding oscillation periods. The levels for global $2.5\sigma$~fluctuations in both directions are indicated by horizontal red lines, i.e., there is a $1.2\%$~probability in the background model that at least one value of~$\Delta q_m$ will lie outside the region bounded by the red lines. \label{fig:deltaqm}}
\end{figure}

To check for consistency with the background model, we consider the test statistic~$\Delta q_m$ (\secref{sec:bkgConsistencyTest}), which is plotted in \figref{fig:deltaqm} for real data from the 2012 observing season of the \emph{Keck Array}.
Since there are roughly $10^4$~frequencies included in our analysis, we use the test statistic~$\Delta \hat{q}$ (\eqref{eq:deltaqhatDef}), which is simply the maximum value of~$\Delta q_m$, to estimate a global $p$-value~$\hat{p}$. In \figref{fig:deltaqm}, we show the $2.5\sigma$~levels for~$\Delta \hat{q}$, and we see that the entire spectrum lies within this region. We find $\hat{p} = 0.14$, which indicates a $1.1\sigma$~signal-like fluctuation in~$\Delta \hat{q}$. As the statistical significance of this fluctuation is far below the threshold set in \secref{sec:unblinding}, we claim no evidence for tension with the background model. 

%%%
\subsection{Upper limits \label{sec:ULresults}}

\begin{figure*}
	\includegraphics[width = \textwidth]{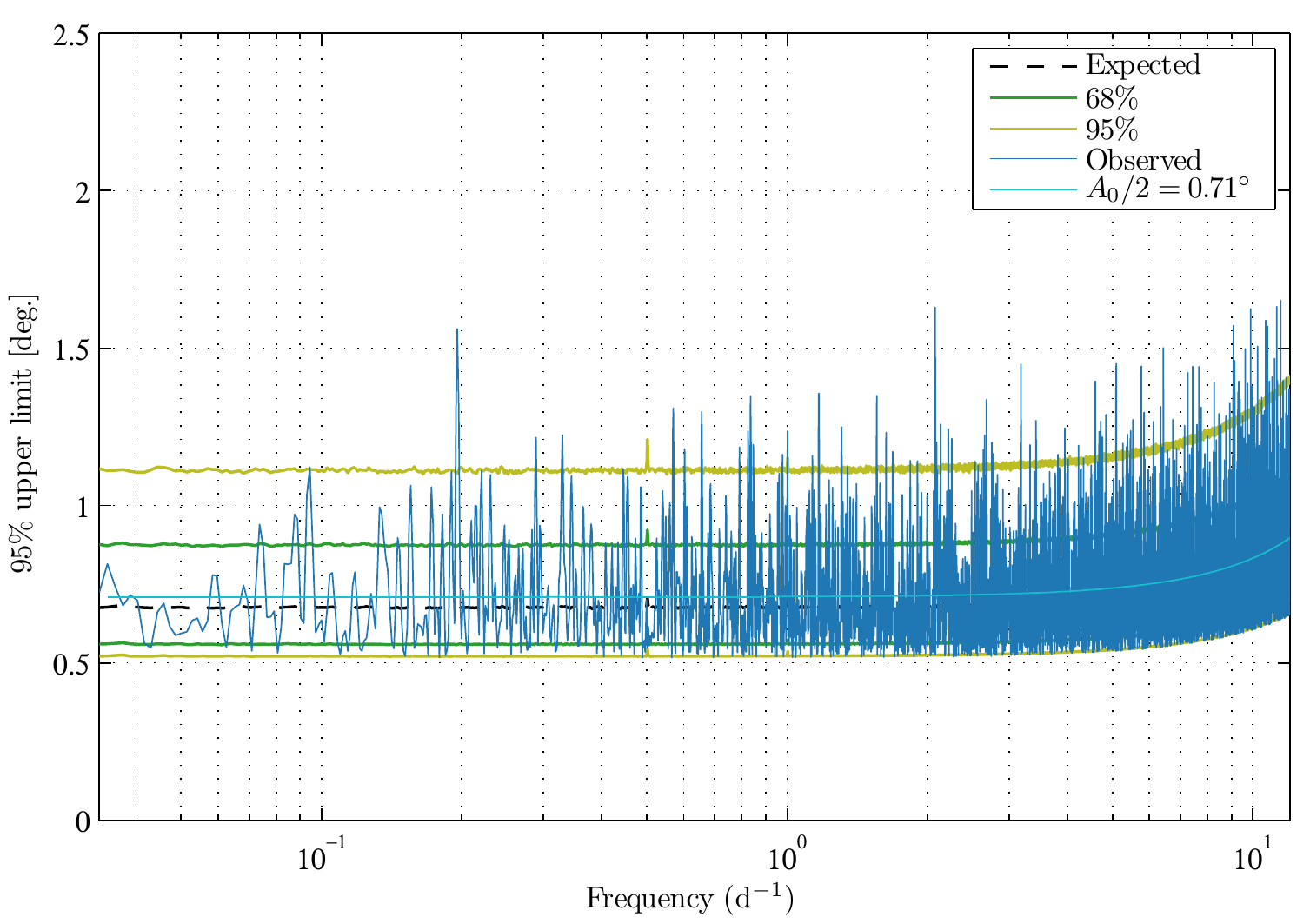}
	\caption{Bayesian $95\%$-confidence upper limits on rotation amplitude~$A/2$ (\secref{sec:BayesianUpperLimits}). We also provide the median expectation (black dashed) from background-only simulations as well as $1\sigma$~(green) and $2\sigma$~(yellow) regions. These expectations represent local rather than global percentiles. With nearly $10^4$~frequencies under consideration, we expect several values outside of the $2\sigma$~region. The question of background consistency is addressed by \figref{fig:deltaqm} and the test statistic~$\Delta \hat{q}$ (\eqref{eq:deltaqhatDef}). The median limit for oscillation periods larger than~$24~\mathrm{hr}$ (frequency less than~$1~\mathrm{d}^{-1}$) is~$0.68^\circ$. For shorter periods (larger frequencies), the limits are degraded due to binning observations in $\sim 1$-$\mathrm{hr}$ scansets (\secref{sec:timeBinning}). Additionally, we plot a smoothed approximation to our upper limits (\eqref{eq:ulmApproxFit}) in cyan. \label{fig:ulm} }
\end{figure*}

The direct observable in this analysis is the Stokes mixing amplitude~$A$. For $A \ll 1$, which is a good approximation in this case, the amplitude of polarization rotations on the sky is~$A/2$. Following the convention of~\cite{Fedderke2019}, we express our upper limits in terms of the rotation amplitude~$A/2$. We follow the prescription of \secref{sec:BayesianUpperLimits} to compute $95\%$-confidence upper limits, and we present the results in \figref{fig:ulm}. 
For oscillation periods longer than one day ($m/(2\pi) < 1~\mathrm{d}^{-1}$), the median limit is
\eq{
A/2 < 0.68^\circ . \label{eq:medianLimit > 24 hr} }
For visual comparison, we also show the expected distribution of upper limits as implied by background-only simulations.

For periods shorter than one day ($m/(2\pi) > 1~\mathrm{d}^{-1}$), the limits degrade by~$\sim 20\%$ as we approach the binning timescale (\secref{sec:timeBinning}). Over the entire frequency range, we can obtain a smoothed approximation to our upper limits by performing a least-squares fit to 
\eq{
	\frac{A}{2} < \frac{A_0}{2 \operatorname{sinc}\parens{ m \Delta t / 2} }  \label{eq:ulmApproxFit}
	}
with $\Delta t = 44.2~\mathrm{min.}$, which is the median scanset duration, and $A_0$~as a free parameter. The sum over $m \not = 0$ of squared residuals is minimized with $A_0/2 = 0.71^\circ$.

\begin{figure*}
	\includegraphics[width = \textwidth]{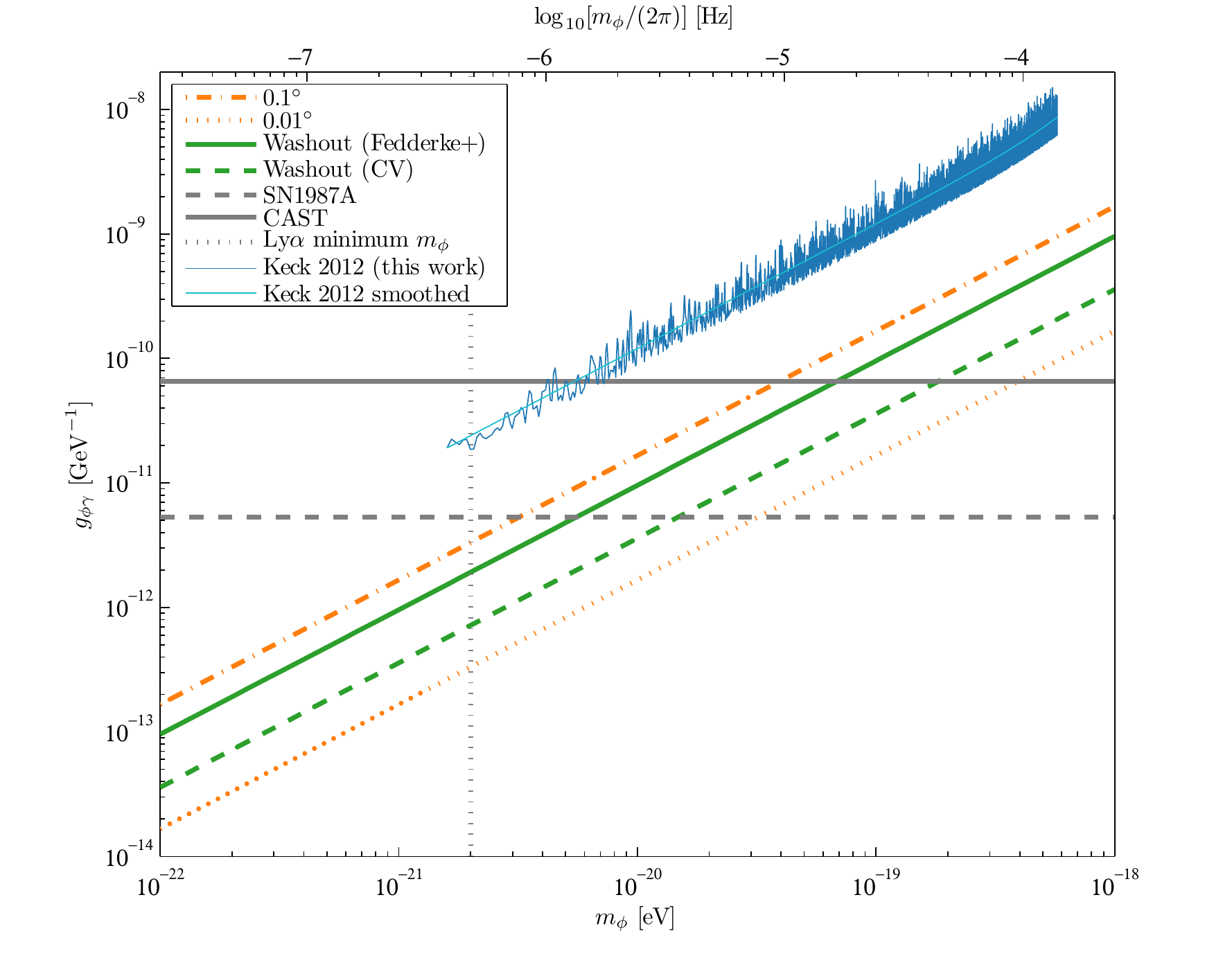}
	\caption{Excluded regions in the mass-coupling parameter space for axion-like dark matter (cf. Fig.~3 in~\cite{Fedderke2019}). All constraints push the allowed regions to larger masses and smaller coupling constants, i.e., toward the bottom right of the figure. If the dark matter is assumed to consist entirely of axion-like particles, i.e., if $\kappa = 1$, then our constraints (blue) are immediately implied by \eqref{eq:phi0} and the results of \figref{fig:ulm}. A smoothed approximation is shown in cyan (\eqref{eq:gLimitApproxFit}). The orange dot-dashed and dotted lines show the constraints that would be achieved if the rotation amplitude were constrained to~$0.1^\circ$ and~$0.01^\circ$, respectively. The green solid line shows the constraint set by Fedderke et al.~\cite{Fedderke2019} by searching for the washout effect (\secref{sec:intro}) in publicly available \emph{Planck} power spectra. The dashed green line shows the cosmic-variance limit for the washout effect. The dashed grey horizontal line shows the limit from searching for a gamma-ray excess from SN1987A~\cite{Payez2015}. The solid grey horizontal line is the limit set by the CAST experiment~\cite{CAST2017}. The dotted grey vertical line is a constraint on the minimum axion mass from observations of small-scale structure in the Lyman-$\alpha$ forest~\cite{Irsic2017}, though we note that several similar bounds have also been set by other considerations of small-scale structure~\cite{Nadler2019,Schutz2020}. \label{fig:exclusion} }
\end{figure*}

To convert our limits on rotation amplitude to the axion parameter space, we identify
\eq{ A = g_{\phi\gamma} \phi_0 \label{eq:ArelatedTogAndPhi0} }
from Eqs.~\ref{eq:f(t)functionalform} and~\ref{eq:f(t)parameterization}. The $m$-dependence of the axion field strength~$\phi_0$ (\eqref{eq:phi0}) implies that our limits on the coupling constant will roughly follow $g_{\phi\gamma} \propto m$. 
In \figref{fig:exclusion}, we present our constraints on the parameter space of axion-like particles from the 2012 observation season of the \emph{Keck Array}.
Combining Eqs.~\ref{eq:phi0}, \ref{eq:ulmApproxFit} and~\ref{eq:ArelatedTogAndPhi0}, we can approximate our limits on the coupling constant by
\spliteq{ g_{\phi\gamma} & < \parens{1.2 \times 10^{-11}~\mathrm{GeV}^{-1}} \operatorname{sinc}^{-1}\parens{ \frac{m}{5.0\times10^{-19}~\mathrm{eV}} } \\
	& \quad\quad \times \parens{ \frac{m}{10^{-21}~\mathrm{eV}} } \parens{ \frac{\kappa \rho_0}{0.3~\mathrm{GeV}/\mathrm{cm}^3}}^{-1/2} . \label{eq:gLimitApproxFit}
}
For periods greater than $24~\mathrm{hr}$, which corresponds to $m < 4.8 \times 10^{-20}~\mathrm{eV}$, we can convert the median limit from \eqref{eq:medianLimit > 24 hr} to
\spliteq{
g_{\phi\gamma} & < \parens{ 1.1 \times 10^{-11}~\mathrm{GeV}^{-1} } \parens{ \frac{m}{10^{-21}~\mathrm{eV}} } \\
& \quad\quad \times  \parens{ \frac{\kappa \rho_0}{0.3~\mathrm{GeV}/\mathrm{cm}^3}}^{-1/2} . 
}
For comparison, we include in \figref{fig:exclusion} the constraints from other probes. Our constraints from only 2012 data do not exclude new regions of parameter space, but we note that the time-domain polarization-oscillation observable is distinct from all others and, consequently, subject to a different set of possible systematic biases. Furthermore, we emphasize that the 2012 observing season of the \emph{Keck Array} represents only a small subset of the total CMB data collected to date and that more sensitive observations will be conducted in the future.

%%%
\section{Conclusions and outlook \label{sec:conclusions}}

We have presented a method to search for axion-like polarization oscillations in the CMB, and we have demonstrated the use of this method with data from the 2012 observing season of the \emph{Keck Array}. The search is compatible with the design and operation of experiments targeting primordial $B$-modes and can be continued by current and future projects with no change to scan strategy nor to low-level data processing.

With only 2012 data from the \emph{Keck Array}, we do not exclude any new regions of the parameter space. We note, however, that we have analyzed only a relatively small fraction of the total BICEP dataset. The \emph{Keck Array} observed for eight seasons, and we have in this work analyzed only one season. Additionally, BICEP3 has been observing at $95~\mathrm{GHz}$ since 2015 with more than twice the mapping speed of the entire \emph{Keck Array}~\cite{Kang2018}. The full BICEP dataset has a survey weight more than an order of magnitude greater than that of the 2012 season.

When more of the BICEP dataset is included in an axion-oscillation analysis, we expect improvements in sensitivity for two reasons. The first is a decrease in residual map noise in the template maps~$\Qcoadd{}$ and~$\Ucoadd{}$. With a better template, we more efficiently extract an oscillation-like signal from the pairmaps. Preliminary investigations indicate that the elimination of residual map noise can improve the per-scanset signal-to-noise ratio by~$\sim 15\%$ for the $150$-$\mathrm{GHz}$ observations analyzed in this work. For frequencies above $\sim 200~\mathrm{GHz}$, which tend to be significantly noisier due to stronger atmospheric fluctuations, the signal-to-noise ratio can be improved by more than a factor of~$2$ by using lower-frequency maps as the CMB templates.
The second improvement in sensitivity will come from the increased sample size. We have verified through simulations that, when the template maps are held constant, our expected upper limits scale approximately as~$1/\sqrt{n}$, where $n$~is the number of scansets included in the analysis. With existing BICEP data, we conservatively anticipate an improvement in upper limits by at least a factor of~3.

Current and future BICEP observations will allow for even more sensitive measurements. We are continuing observations with BICEP3 at $95~\mathrm{GHz}$. The \emph{BICEP Array} has begun a staged deployment of four new receivers of similar size to BICEP3~\cite{Schillaci2020}. The first receiver, which observes at~$30$ and~$40~\mathrm{GHz}$, achieved first light in February~2020. The second and third receivers will observe at~$150$ and~$95~\mathrm{GHz}$, respectively, and the fourth receiver will observe at~$220$ and~$270~\mathrm{GHz}$. 

Additional improvements in sensitivity can be achieved by correlating in the time domain with other CMB experiments. The South Pole Observatory is a formal partnership between the BICEP collaboration and the South Pole Telescope (SPT) collaboration. The current generation of SPT, which is called SPT-3G~\cite{Bender2018}, has been observing from the South Pole since 2017. While the BICEP dataset has greater integrated polarization sensitivity, SPT has greater angular resolution and is, therefore, sensitive to more polarization modes. All else being equal, a higher-resolution CMB experiment is more sensitive to polarization oscillations due to the increased number of modes, though this advantage is less significant for multipoles larger than $\ell \sim 2000$, where the CMB anisotropies are suppressed.

The CMB Stage-4 (CMB-S4) project will contain more than an order of magnitude more detectors than any current-generation experiment, and this will provide yet another boost in sensitivity~\cite{Carlstrom2019,Abazajian2019}. An axion-oscillation search imposes few requirements on the design and scan strategy of CMB-S4, since the main elements are nothing more than sensitive, repetitive measurements of CMB polarization. The search is more sensitive at CMB-dominated frequency bands like $95$ and $150~\mathrm{GHz}$, since the global oscillation affects only the CMB component of the polarization field. To take full advantage of the polarization information in the CMB and thereby increase the signal-to-noise ratio, higher-resolution instruments are preferred, e.g., with aperture diameters of $5$-$10~\mathrm{m}$, which allow for sensitivity to polarization modes into the CMB damping tail.

The methods presented in this work can be adapted with relatively minor alterations to analyze data from other CMB polarimetry experiments. Some of our analysis choices take advantage of unique characteristics of the \emph{Keck Array}, and we have attempted to draw attention to those experiment-specific assumptions. As the signal is coherent over large time and length scales (\secref{sec:intro}), observations from several CMB experiments can be combined to protect against systematics and improve sensitivity.

\begin{acknowledgments}
We thank Adam~J. Anderson, Aviv~R. Cukierman, Michael~A. Fedderke and Ethan~O. Nadler for useful conversations.
The BICEP/\emph{Keck Array} projects have been made possible through a series of grants from the National Science Foundation including 0742818, 0742592, 1044978, 1110087, 1145172, 1145143, 1145248, 1639040, 1638957, 1638978, 1638970 \& 1836010 and by the Keck Foundation.
The development of antenna-coupled detector technology was supported
by the JPL Research and Technology Development Fund and Grants No.\
06-ARPA206-0040 and 10-SAT10-0017 from the NASA APRA and SAT programs.
The development and testing of focal planes were supported
by the Gordon and Betty Moore Foundation at Caltech.
Readout electronics were supported by a Canada Foundation
for Innovation grant to UBC.
The computations in this paper were run on the Odyssey cluster
supported by the FAS Science Division Research Computing Group at
Harvard University.
The analysis effort at Stanford and SLAC was partially supported by the Department of Energy, Contract DE-AC02-76SF00515.
We thank the staff of the U.S. Antarctic Program and in particular
the South Pole Station without whose help this research would not
have been possible.
Most special thanks go to our heroic winter-overs Robert Schwarz
and Steffen Richter.
We thank all those who have contributed past efforts to the BICEP/\emph{Keck Array}
series of experiments, including the BICEP1 team.
\end{acknowledgments}

\bibliography{axionOscillations_Keck2012_PRD}

\end{document}